\documentclass[reprint,aps,prb,superscriptaddress,nofootinbib]{revtex4-2}

\usepackage{graphicx,amssymb,subcaption}

\usepackage{dsfont,mathrsfs,xcolor,url,verbatim,booktabs}
\usepackage{centernot}

\usepackage{hyperref}
\hypersetup{colorlinks=true,allcolors=black}
\usepackage{hypcap}
\usepackage{bookmark}
\usepackage{tikz}
\usetikzlibrary{patterns.meta}

\usepackage{amsmath, braket, array, multirow, float, empheq, tabstackengine, cases}
\setstackgap{L}{1.2\normalbaselineskip}
\setstacktabbedgap{.4em}
\fixTABwidth{T}

\DeclareMathOperator{\Tr}{Tr}

\newcommand\ignore[1]{{}}
\newcommand{\Hinf}{H}%alternative: {H_\text{I}}
\newcommand{\Hsi}{H_\text{SI}}
\newcommand{\beq}{\begin{equation}}
\newcommand{\eeq}{\end{equation}}
\newlength{\matrixcolwidth}%used in a couple of places to make all columns of matrix the same width
\newif\ifmarked
\markedfalse   % version annotée ; mettre \markedfalse pour la version propre
\ifmarked\else\colorlet{red}{black}\fi

\begin{document}

\preprint{Bulk-Boundary correspondence Iakoub}

\title{Bulk-Boundary Correspondence in Semi-Infinite Chains from
Sublattice Zeros}
\author{Ilya Iakoub}
\email{ilya.iakoub@umontreal.ca}
\affiliation{Département de physique, Université de Montréal, Montréal, QC, Canada, H3C 3J7}
\author{Nicolas Levasseur}
\email{nicolas.levasseur@umontreal.ca}
\affiliation{Département de physique, Université de Montréal, Montréal, QC, Canada, H3C 3J7}
\author{Kylian Lionnet}
\email{kylian.lionnet@umontreal.ca}
\affiliation{Département de physique, Université de Montréal, Montréal, QC, Canada, H3C 3J7}
\author{Richard MacKenzie}
\email{richard.mackenzie@umontreal.ca}
\affiliation{Département de physique, Université de Montréal, Montréal, QC, Canada, H3C 3J7}

\begin{abstract}

We provide an alternative derivation of the bulk-boundary correspondence for semi-infinite chains. To describe edge states, we analytically continue the usual Bloch Hamiltonian to complex wave vectors $k$. To start, we note that the zeros of a Bloch wavefunction are related to edge states, at least in systems with nearest-neighbour hoppings. We show that an analytically continued Bloch Hamiltonian with chiral symmetry has exceptional points, where two different states coalesce. These special points are related to the adiabatic protection of edge states. {\color{red}%
We derive a winding number that counts the exceptional states which, together with the boundary conditions, give a lower bound on the number of edge states protected by chiral symmetry. This bound is saturated in the models we consider.
}%end red

\end{abstract}

\maketitle

\section{Introduction}
\label{sec:intro}

{\color{red}%
The bulk-boundary correspondence is one of the central ideas in the theory of topological insulators: a topological invariant computed from the Bloch Hamiltonian of an infinite, translation-invariant system determines the number of protected states localized at the boundary of a finite or semi-infinite one~\cite{Hatsugai, Hatsugai_Riemann, Ryu_Hatsugai, Ryu_2010, tr_Q, Asboth}. The simplest setting in which it can be studied is that of one-dimensional chains with chiral (sublattice) symmetry, of which the Su-Schrieffer-Heeger (SSH) model~\cite{SSH} is perhaps the prototypical example. There, the relevant invariant is a winding number, or equivalently the Zak phase~\cite{Zak}, and the protected boundary states are zero-energy edge states. Rigorous proofs of the correspondence exist~\cite{K-Theory, Toeplitz_algebra}; they rely on mathematical tools, such as $K$-theory and Toeplitz index theory. While these tools are powerful and general, they do not always make the underlying physical mechanism apparent.

Even in one dimension, the correspondence has some well-known subtleties. The winding number and the Zak phase depend on the choice of unit cell (and, for the Zak phase, of gauge), and they correctly predict the number of edge states only when the unit cell is chosen to be compatible with the termination of the chain~\cite{Delplace2011, Rhim2018, Asboth}. Furthermore, chains with more than two sites per unit cell or with longer-range hoppings, such as the trimer SSH model~\cite{SSH3} or SSH models with longer-range couplings~\cite{Zeros_of_order_2}, may have edge states that are not protected by any symmetry, or several protected edge states at a single edge. Refinements of the Zak phase restricted to a single sublattice have been shown to predict the number of edge states in several such cases~\cite{SZP, SZP2, SSH3}.

A natural way to make the connection between bulk and boundary explicit is to continue the Bloch Hamiltonian to complex wave vectors $k$, or equivalently to replace $e^{ika}$ by a complex variable $z$. States with $|z|<1$ are not acceptable in an infinite chain, but they decay away from the edge of a semi-infinite one, and are therefore the natural candidates for edge states. This idea goes back to Kohn's study of the analytic properties of Bloch waves~\cite{Kohn1959}. It underlies Hatsugai's treatment of edge states in the quantum Hall effect, in which edge states are associated with zeros of Bloch functions on a Riemann surface~\cite{Hatsugai_Riemann, Hatsugai}, as well as the analysis of edge states in Dirac Hamiltonians by Mong and Shivamoggi~\cite{BB_Chern}. It also lies at the heart of the generalized transfer matrices of Dwivedi and Chua~\cite{Dwivedi_Chua} and of the generalization of Bloch's theorem to arbitrary boundary conditions of Alase \emph{et al.}~\cite{Alase2017}, in which the eigenvalue problem of a finite or semi-infinite system is split into bulk equations, solved by states with complex crystal momenta, and boundary equations that select linear combinations of these. For chiral chains, Balabanov \emph{et al.}~\cite{Balabanov2021} expressed the winding number as the number of zeros minus the number of poles, inside the unit disk, of the determinant of the analytically continued off-diagonal block of the Bloch Hamiltonian, and used this to construct the edge modes explicitly, including at critical points where the bulk gap closes. The same analytic continuation is also central to the non-Bloch band theory of non-Hermitian systems, where it is used to construct the generalized Brillouin zone that replaces the unit circle~\cite{Yao_Wang, Yokomizo_Murakami, Zhang_Yang_Fang, Yang_Zhang_Fang_Hu}. In the Hermitian systems considered here, the Brillouin zone remains the unit circle, and the interior of the unit disk serves only to host boundary solutions.

Once $k$ is complex, the Bloch Hamiltonian is no longer Hermitian, and it can have exceptional points: values of $z$ at which it is not diagonalizable and two (or more) eigenstates coalesce~\cite{Nimrod, Physics_of_E.P., Topology_of_exceptional_points, non-Hermitian_lit_review}. Exceptional points have been related to edge states in several contexts. Lee~\cite{Lee2016} showed that a non-Hermitian chain whose Bloch Hamiltonian encircles an exceptional point has a fractional winding number and a single, chirally protected, zero-energy edge state whose existence is due to the Hamiltonian being defective. Closest in spirit to the present work, Mandal~\cite{mandal_exceptional_2015} argued that Majorana zero modes in chiral topological superconductors are associated with exceptional points of the analytically continued Bloch Hamiltonian, noting that such exceptional points appear in the presence of chiral symmetry.} \textcolor{red}{Here, by contrast, we consider general chiral lattice chains with explicit terminations, and use the exceptional points to count the protected edge states at a given termination, including when several boundary conditions must be satisfied.}

{\color{red} In this paper, we provide an alternative derivation of the bulk-boundary correspondence for semi-infinite chains with chiral symmetry, with $N$ sites per unit cell and hoppings of arbitrary (finite) range. The derivation relies on complex analysis and linear algebra rather than on $K$-theory or index theorems. Our aim is to make the mechanism behind the correspondence as transparent as possible, and to obtain a counting of edge states that does not depend on the choice of unit cell or gauge. The analysis rests on two observations. First, under certain conditions (in particular for nearest-neighbour hoppings across the boundary), an edge state is an analytically continued Bloch state with $|z|<1$ whose amplitude vanishes on one sublattice, namely the one just outside the boundary; edge states can therefore be counted by counting such sublattice zeros inside the unit disk, using the argument principle. Since the domain of integration excludes $z=0$, where a change of unit cell only modifies the order of a pole, this count is independent of the choice of unit cell, unlike the winding number or the Zak phase. The resulting expression is closely related to the sublattice Zak phase~\cite{SZP, SZP2, SSH3}. Second, in chiral-symmetric chains the zero-energy Bloch states that vanish on one sublattice occur at exceptional points of the analytically continued Bloch Hamiltonian. These states can only cross the unit circle, and therefore disappear from the semi-infinite chain, when the bulk gap closes, which provides a simple picture of their adiabatic protection. Counting them leads to a winding number which, when the corresponding boundary condition is sufficient, gives the number of protected edge states, and which reduces to the usual chiral winding number~\cite{tr_Q, Chiral_winding} when the two sublattices are combined.

Our counting of zero-energy states is closely related to that of Balabanov \emph{et al.}~\cite{Balabanov2021}; our emphasis is on the interpretation of these states as sublattice zeros of Bloch states and as exceptional points of the analytically continued Bloch Hamiltonian, and on the role of the boundary. While Mandal~\cite{mandal_exceptional_2015} reached a conclusion similar to ours regarding exceptional points, that work concerns Majorana fermions and its results are derived in a significantly different manner; in particular, its counting formula is based on sign changes relative to a reference trivial phase, whereas we derive a winding number that counts the exceptional states on each sublattice absolutely, and treat lattice chains with explicit terminations. For longer-range hoppings, more than one boundary condition must be satisfied. Following the general strategy of Refs.~\cite{Alase2017, Dwivedi_Chua}, we determine the number of independent boundary conditions for a given termination of the chain and show how they combine with the number of exceptional states to give a lower bound on the number of protected edge states.

The paper is organized as follows. \autoref{sec:Bloch} covers some prerequisites for the rest of the paper and discusses the analytic continuation of the Bloch Hamiltonian and its para-Hermiticity. In \autoref{sec:bulkBoundaryStates}, we relate states of semi-infinite chains to those of the corresponding infinite chain. In \autoref{sec:countingEdge}, we derive a winding number that counts (not necessarily adiabatically protected) edge states. In \autoref{sec:adiabaticProtection}, we establish the conditions under which the number of edge states is an adiabatic invariant, relate this to the exceptional points of the analytically continued Bloch Hamiltonian, and treat the case of longer-range hoppings across the boundary. In \autoref{sec:extendedSSH}, we illustrate our results on an extended SSH model with longer-range hoppings. \autoref{sec:conclusion} concludes the paper with some closing remarks. The boundary conditions of semi-infinite chains are analyzed in App.~\ref{app:boundarySemiInfinite}, and a four-band example is treated in App.~\ref{app:4band}.
}%end of red

\section{Analytic Continuation of Bloch Hamiltonian}
\label{sec:Bloch}

Our goal is to find edge states of a semi-infinite chain with $N$-site unit cell and hoppings which are $N$-site periodic but which are not necessarily limited to nearest neighbours. Since a semi-infinite chain is essentially a truncated infinite chain plus one or more boundary conditions, we begin our search by briefly looking at the states of the corresponding infinite chain, with Hamiltonian $\Hinf$. We will then transition to the semi-infinite chain by first relaxing the normalization condition on eigenstates and then incorporating the appropriate boundary conditions.

Let the sites be labelled by the unit cell index $p \in \mathbb{Z}$, and the sublattice index $\mu \in \{1,2,...,N\}$, so $|p,\mu\rangle$ represents a particle localized on the $\mu^\text{th}$ site of the $p^\text{th}$ unit cell.
Periodicity implies that $\langle p+m, \mu'| \Hinf |p,\mu\rangle$ is independent of $p$, so we write 
\begin{equation}\label{eq:hoppings}
    \langle p+m,\mu'|\Hinf|p,\mu\rangle \equiv t^m_{\mu',\mu} = (t^{-m}_{\mu,\mu'})^*,
\end{equation}
where the second equality follows from the Hermiticity of $\Hinf$. The matrix $t^m$ and its Hermitian conjugate contain the $m^\text{th}$ inter-cell hopping parameters.

The Hamiltonian can be viewed as an infinite-dimensional matrix of $(N\times N)$-dimensional blocks. The blocks along the diagonal contain the matrix $t^0$; those along the first subdiagonal below and above contain $t^1$ and $(t^1)^\dagger$, respectively; and so on. In what follows, it will be useful to define $M$, the farthest hopping in terms of unit cells; thus, beyond the $M^\text{th}$ subdiagonal above and below, $\Hinf=0$. We also define the hopping range $R$ as follows: if only nearest-neighbour hoppings are present, $R=1$; if next-nearest-neighbour hoppings are also present, $R=2$; etc.

Bloch states are formed via Fourier transformation
\begin{equation}\label{eq:Bloch state}
    |k;\mu\rangle \equiv \sum_{p} e^{ika p} |p,\mu\rangle,
\end{equation}
where $k \in (-\pi/a,\pi/a]$ and $a$ is the lattice constant. (Note the semicolon on the left and comma on the right, signifying that the first state label is a wave number on the left and a unit cell index on the right. This convention will be followed throughout the paper.) These states obey the following orthonormality condition:
\begin{equation}
    \langle k';\mu'|k;\mu\rangle = 2\pi \delta(k-k') \delta_{\mu',\mu},
\end{equation}
with $\delta(k-k')$ and $\delta_{\mu',\mu}$ being the Dirac and Kronecker deltas, respectively. The $(N\times N)$-dimensional Bloch Hamiltonian $h_k$ is defined by
\begin{align}
    \langle k';\mu'|\Hinf|k;\mu\rangle &\equiv 2\pi \delta(k-k')\langle \mu'|h_k|\mu\rangle \nonumber\\
    &= 2\pi \delta(k-k') (h_k)_{\mu',\mu}
\end{align}
giving
\begin{equation}\label{eq:Bloch matrix element}
    (h_k)_{\mu',\mu}=\sum_{m=-M}^{M}e^{-imka}t_{\mu',\mu}^m.
\end{equation}
Hermiticity of $h_k$ (for $k$ real) is guaranteed by \eqref{eq:hoppings}.

For an infinite chain, states with $k \notin \mathbb{R}$ are unacceptable since they are non-normalizable: they diverge exponentially as $p\to \pm \infty$ for $\Im{(k)} \lessgtr 0$. However, for a semi-infinite or finite chain, such states are not only permitted — they are of great interest. Since they grow exponentially towards the boundary, they are perfect candidates for edge states. To accommodate these states we must replace $e^{ika}\to z \in \mathbb{C}$, so in analogy with \eqref{eq:Bloch state} and \eqref{eq:Bloch matrix element} we define
\begin{align}
    |z;\mu\rangle &\equiv \sum_p z^p |p,\mu\rangle,\\
    h(z)_{\mu',\mu} &\equiv \sum_{m=-M}^{M}z^{-m}t^m_{\mu',\mu}.\label{eq:Analytic continuations}
\end{align}
This gives rise to various complications since the analytically continued Bloch Hamiltonian \eqref{eq:Analytic continuations}, viewed as a function of the complex parameter $z$, is only Hermitian on the unit circle $|z|=1$. Thus, we lose many properties normally taken for granted such as orthonormality of eigenstates, diagonalizability by a unitary matrix, etc.

Fortunately, $h(z)$, while not Hermitian for $|z|\neq1$, has a related property (with related consequences) known as para-Hermiticity~\cite{Para_Hermitian,Para_Hermitian2}:
\begin{equation}
    h(z)=h(1/z^*)^\dagger.
\end{equation}
(On the unit circle $1/z^* = z$, so we recover the usual Hermiticity.) The Schrödinger equation for $h(z)$ is 
\begin{equation}\label{eq:dog1}
    h(z)|\alpha^{(m)}(z)\rangle = E^{(m)}(z)|\alpha^{(m)}(z)\rangle,
\end{equation}
where $(m)$ is the band index and $|\alpha^{(m)}\rangle$ is an $N$-component eigenvector.

Para-Hermiticity has the following consequences for \eqref{eq:dog1}:
\begin{enumerate}
    \item The energies satisfy
    \begin{equation}
        E^{(m)}(z)=(E^{(m)}(1/z^*))^*
        \label{eq:EnotReal}
    \end{equation}
    (so $E^{(m)}$ is not, in general, real).
    \item The Bloch eigenstates satisfy
    \begin{equation}\label{eq:scalar product}
        \bra{\alpha^{(m)}(1/z^*)}\alpha^{(n)}(z)\rangle = \mathcal{N}^{(m)}(z) \delta_{m,n},
    \end{equation}
    where $m,n$ are band indices and $\mathcal{N}^{(m)}(z) \in \mathbb{C}$.
\end{enumerate}
The second property implies that, whenever $h(z)$ has a complete basis of eigenstates (which, as we shall see, is not always the case), it can be diagonalized by a para-unitary matrix $U(z)$, that is, one satisfying $U^{-1}(z)=U^\dagger(1/z^*)$.

Importantly for the following, $h(z)$ is analytic for $z$ nonzero and finite, as can be seen from \eqref{eq:Analytic continuations}. Therefore, there is always a choice of normalization and gauge for which $\ket{\alpha^{(m)}(z)}$ is nonzero and analytic, except for possible branch points~\cite{Kato}; these arise in both $E^{(m)}(z)$ and $\ket{\alpha^{(m)}(z)}$ at non-diagonalizable (exceptional) points of $h(z)$~\cite{Nimrod, Topology_of_exceptional_points, Physics_of_E.P., Higher_order_E.P.}. Near an exceptional point, both $E^{(m)}(z)$ and $\ket{\alpha^{(m)}(z)}$ exhibit branch-point behaviour with a local dependence of the form $(z-z_0)^{1/r}$, where the order $r$ depends on the structure of the Hamiltonian.

\section{Bulk and boundary states}
\label{sec:bulkBoundaryStates}

Energy eigenstates for the infinite chain are easily constructed in terms of the Bloch eigenstates. We define $\ket{p,\alpha^{(m)}(z)}$ to be the Bloch eigenstate $\ket{\alpha^{(m)}(z)}$ in the $p^\text{th}$ unit cell and zero outside; then the eigenstates (ignoring normalizability) are
\[
\ket{\Psi^{(m)}(z)} \equiv \sum_p z^p \ket{p,\alpha^{(m)}(z)}.
\]
Given that inter- and intra-band degeneracies are possible, the most general eigenstate of energy $E$ is
\begin{equation}\label{eq:arbitrary state}
    \ket{\Psi(E)}=\sum_{(m,z) \in g(E)} C_{m,z}\ket{\Psi^{(m)}(z)},
\end{equation}
where $C_{m,z}$ are complex coefficients and $g(E)$ is the set of $(m,z)$ values for which the energy is $E$ (that is,
$g(E) \equiv \{(m,z): E^{(m)}(z) = E\}$).

Consider now a semi-infinite chain (henceforth referred to as the system) with edge on the left and extending to infinity on the right. (The analysis can easily be adapted to a semi-infinite chain extending to infinity on the left.) Let $\Hsi$ denote the Hamiltonian of the system. Any state of the infinite chain corresponds to a state of the system by \emph{reduction} (dropping the components of the state corresponding to the excised sites); we write
\beq\label{eq:restriction}
\ket{\psi} \equiv {\cal R} \ket{\Psi}.
\eeq
States with $|z|=1$ are bulk states; states with $|z|<1$ are edge states; states with $|z|>1$ grow exponentially and are eliminated by normalization.

Applying ${\cal R}$ to the energy eigenstate \eqref{eq:arbitrary state} will not automatically result in an eigenstate of the system; the coefficients $\{C_{m,z}\}$ must satisfy certain boundary conditions that depend on the details of the model. Consider the case where one boundary condition is $\bra{p_b,\mu_b}\Psi(E)\rangle=0$, where $\ket{p_b,\mu_b}$ is the first site exterior to the chain. Although other boundary conditions may be required (see App.~\ref{app:boundarySemiInfinite} for a detailed discussion), it will give us important insights, and we shall focus on this specific condition throughout the next sections.

We will assume that, for edge states, the reduction of \eqref{eq:arbitrary state} satisfies the boundary condition with a single nonzero coefficient. In other words, we will assume that the edge states can be obtained from  a single analytically continued Bloch state.
{\color{red}%
This is appropriate when the single boundary condition $\bra{p_b,\mu_b}\Psi(E)\rangle=0$ is sufficient, in particular when the only hoppings across the boundary are to nearest neighbours (see App.~\ref{app:boundarySemiInfinite}). For longer-range hoppings, several boundary conditions must be satisfied, and edge states are in general linear combinations of several analytically continued Bloch states; this case is treated in \autoref{subsec:movingAway}.
}%end red
Under this assumption, edge states are of the form
\begin{equation}
\label{eq:dog2}
    \ket{\psi_\text{edge}}={\cal R}\ket{\Psi^{(m)}(z)}
\end{equation}
for one or more $(m,z)\in g(E)$.
The boundary condition implies that
\begin{align}\label{eq:edge state boundary condition}
    \langle p_b, \mu_b\ket{\Psi^{(m)}(z)}&=\bra{p_b,\mu_b}\sum_{p}z^p \ket{p,\alpha^{(m)}(z)}\nonumber\\& \propto \bra{\mu_b}\alpha^{(m)}(z)\rangle \equiv \alpha_b^{(m)}(z) =0,
\end{align}
where we have defined $\alpha_b^{(m)}(z)$, the Bloch state amplitude on the sublattice $\mu_b$ associated with the state \eqref{eq:dog2}. Thus, for an edge state of this form the wave function at every $N$-th site along the chain vanishes.

We show in \autoref{sec:chiralSymmetry} that the number of (analytically continued) Bloch states satisfying \eqref{eq:edge state boundary condition} can be adiabatically protected under certain conditions. If these conditions are satisfied, it suffices for a system to be adiabatically connected to one where \eqref{eq:edge state boundary condition} holds trivially in order for it to have edge states that can be obtained from a single Bloch state.

\section{Counting Edge States Satisfying the Boundary Condition}
\label{sec:countingEdge}

\subsection{Edge states without branch cuts}
\label{subsec:edgeStatesWithoutBranchCuts}
Assuming that edge states are analytically continued Bloch states with a zero in one sublattice, we can construct a winding number that gives the number of edge states  satisfying the boundary condition $\alpha^{(m)}_b(z)=0$ (assumed to be sufficient throughout this section), as follows. If $h(z)$ is diagonalizable for $z$ nonzero and finite, then $\alpha^{(m)}_b(z)$ can be assumed to be analytic on the same domain~\cite{Kato}. (It was noted at the end of \autoref{sec:Bloch} that $h(z)$ is not always diagonalizable, the implications of which will be explored in detail in \autoref{subsec:exceptionalPoints}.) Let $D$ be a bounded domain in $\mathbb{C}\backslash \{0\}$. Denote the zeros of $\alpha^{(m)}_b(z)$ lying within $D$ and their multiplicities, respectively, by $z_j$ and $m_j$ ($j=1,2,\dots$). Furthermore, assume that no zeros occur on the boundary $\partial D$. Then the argument principle applied to the function $\alpha^{(m)}_b(z)$ gives
\begin{equation}\label{eq:argument principle}
    \frac{1}{2\pi i}\oint_{\partial D} dz \frac{\partial_z\alpha^{(m)}_b}{\alpha^{(m)}_b} = \sum_j m_j.
\end{equation}

\begin{figure}[t]
    \centering
    \includegraphics[width=1\linewidth]{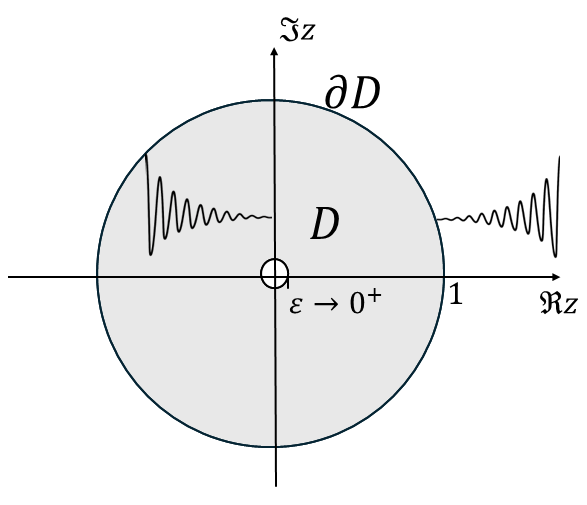}
    \caption{The domain $D$ and its boundary $\partial D$ in the complex $z$ plane. $D$ consists of allowed values of $z$ for a semi-infinite chain with boundary on the left. (Were the boundary on the right, the complementary region ($|z|>1$) would be allowed.) The appearance of states in each region is indicated.}
    \label{fig:DLDR}
\end{figure}

Given that the boundary of the system is on the left, we need only consider states with $0<|z| \leq 1$, of which those with $|z|<1$ are edge states. Thus, we will restrict $D$ to $D = \{0 < |z| < 1\}$, as illustrated in \autoref{fig:DLDR}. The number of edge states coming from the $m$-th band is bounded from above by
\begin{equation}\label{eq:SZP}
    \nu_{b}^{(m)} \equiv \frac{1}{2\pi i}\oint_{\partial D} dz \frac{\partial_z\alpha^{(m)}_b}{\alpha^{(m)}_b}
\end{equation}
where $\partial D$ consists of the unit circle (integrated anticlockwise) and a circle of infinitesimal radius centred at the origin (integrated clockwise). This bound is saturated (that is, $\nu_{b}^{(m)}$ gives the exact number of edge states from the $m$-th band) if $m_j=1$ for all $j$, while if any multiplicities are greater than one \eqref{eq:SZP} will overcount the number of edge states.

Multiplicities greater than one do not occur generically (they require fine tuning of the parameters in the Hamiltonian), and they are unstable against a small perturbation. For example, suppose $\alpha_b(z)=\alpha_0 (z-z_0)^2$, so that $z_0$ is a zero of multiplicity 2 and consider the small perturbation $\alpha_b(z)\to \alpha_b'(z)=\alpha_b(z)+\varepsilon$, where $|\varepsilon| \ll 1$. Now, $\alpha_b'(z)=\alpha_0(z-z_0-i\sqrt{\varepsilon/{\alpha_0}})(z-z_0+i\sqrt{\varepsilon/{\alpha_0}})$, which has two zeros of multiplicity 1. This argument is easily generalized to zeros of higher order. Thus, while \eqref{eq:SZP} gives the \emph{maximum} number of edge states, generically it gives the \emph{exact} number of edge states, and henceforth we will assume this to be true.

We note the similarity between \eqref{eq:SZP} and the previously established sublattice Zak phase~\cite{SZP, SZP2, SSH3}. In fact, if we choose a normalization such that $|\alpha_b|=1$ (as was done in~\cite{SSH3}), then \eqref{eq:SZP} reduces to
\begin{align}
    \nu_{b}^{(m)}&=\frac{1}{2\pi i}\oint_{\partial D} dz (\alpha^{(m)}_b)^*\partial_z\alpha^{(m)}_b\nonumber\\
    &=\frac{1}{2\pi i}\oint_{\partial D} dz \bra{\alpha^{(m)} (z)}P_b\partial_z P_b\ket{\alpha^{(m)}(z)}
    \label{eq:SZP real}
\end{align}
 where $P_b=\ket{\mu_b}\bra{\mu_b}$ is the projector on the boundary sublattice. This is none other than the Zak-Berry phase of the sublattice $\mu_b$ (albeit with a different domain of integration). It has already been shown that the sublattice Zak phase is able to correctly predict the number of edge states in certain cases~\cite{SSH3,SZP,SZP2}. Note that the chosen normalization makes $\ket{\alpha^{(m)}(z)}$ non-analytic. Since it does not affect the argument of $\alpha_b^{(m)}(z)$, it will still give the same quantity as \eqref{eq:SZP} in the non-normalized case. To highlight a side point, any choice of gauge and normalization that keeps $\ket{\alpha^{(m)}(z)}$ analytic must be coherent with the argument principle; thus, \eqref{eq:SZP} is invariant under an analytic choice of gauge and normalization.\footnote{In practice, however, we may not have control over the gauge, for example in a numeric calculation. To remove the reliance on an analytic gauge, it is possible to compute the difference of $\nu^{(m)}_{b}$ for two different boundaries.}
 
 Finally, $\nu_b$ does not depend on the choice of unit cell, since a change of unit cell must be accompanied by a corresponding change of which sublattice amplitude is named ``$\alpha_b$''. In practice, a change of unit cell results in a change of the order of the poles at $z=0,\infty$, but does not affect the multiplicities or positions of the zeros. Therefore, since $D$ excludes $z=0,\infty$, $\nu_b$ remains unchanged. 

\subsection{Edge states with branch cuts}
\label{subsec:edgeStatesBranchCuts}

Of particular interest in what follows will be edge states arising at non-diagonalizable points of $h(z)$ in $z$-space. As has been noted in \autoref{sec:Bloch}, the eigenstates and eigenvalues of $h(z)$ can develop root singularities at the non-diagonalizable, exceptional, points~\cite{Nimrod,Topology_of_exceptional_points}. We will only be concerned with exceptional points corresponding to the coalescence of $r$ eigenstates and eigenvalues~\cite{non-Hermitian_lit_review}. For such states, \eqref{eq:SZP} will return $(\text{total multiplicity})/{r}$, where $r \in \mathbb{N}$ is the order of the root singularity~\cite{Higher_order_E.P.}. For example, in the case of a square root singularity, which is particularly common (for a concrete example, see \autoref{sec:extendedSSH}), $\nu^{(m)}_{b}$ will give half of the total multiplicity of the zeros of $\alpha_b^{(m)}$. In the presence of exceptional points, the index $\nu_{b}^{(m)}$ by itself can only tell us that there are edge states (provided that the boundary condition \eqref{eq:edge state boundary condition} is sufficient), rather than telling us the maximum number of edge states. Because in that case all edge states have the same branch cut, we can compute the number of Bloch states satisfying \eqref{eq:edge state boundary condition} with
\begin{equation}\label{eq:asdf}
    \text{\# e.s.}= \sum_{m=1}^{N}\nu_{b}^{(m)},
\end{equation}
since each band with an edge state will contribute $1/r$, and there are $r$ such bands coalescing.

\section{Adiabatic Invariant}
\label{sec:adiabaticProtection}

Consider the manner in which a continuous change in the Bloch Hamiltonian can affect the number of edge states that satisfy a boundary condition of the form \eqref{eq:edge state boundary condition}. Under a continuous change of the Hamiltonian that does not close the gap (in other words, an adiabatic change), the eigenstates, and thus the coefficients $\alpha^{(m)}_b(z)$, will also change continuously. We wish to show that the zeros of $\alpha_b^{(m)}(z)$ cannot disappear from the complex plane during this evolution for $z$ nonzero and finite (so that $h(z)$, and therefore $\alpha_b^{(m)}(z)$, is analytic).

Assume the evolution of the Hamiltonian is parameterized by a parameter $t\in [0,1]$, with the original Hamiltonian corresponding to $t=0$. Then $\alpha_b^{(m)}(z) \to \alpha_b^{(m)}(z,t)$ with $\alpha_b^{(m)}(z,0) = \alpha_b^{(m)}(z)$. Any zeros of $\alpha_b^{(m)}(z,t)$, denoted $z_0(t)$, are given by
\begin{equation}
    \alpha_b^{(m)}(z_0,t)=0 \implies \frac{d}{dt} \alpha_b^{(m)}(z_0(t),t)=0.
\end{equation}
The function $\alpha_b^{(m)}$ remains analytic throughout the evolution. We can assume that, in the neighbourhood of $z_0$, $\partial_z\alpha_b^{(m)}(z,t) \Big|_{z_0}\neq 0$.\footnote{This assumption is obviously true for zeros of multiplicity 1 since, in the neighbourhood of the zero, $\alpha_b^{(m)}\sim(z-z_0)$. As we have seen, we can easily split a zero of order $n$ into $n$ zeros of order 1 by a small perturbation. The same argument will apply then and the total multiplicity of zeros is conserved.}
By the implicit function theorem there exists a unique continuous path $z_0(t)$ given by the solution of the equation
\[
\frac{d}{dt}z_0 = -\frac{\partial \alpha}{ \partial t}\Big/{\frac{\partial \alpha}{ \partial z}}\Bigg|_{(z_0(t),t)}.
\]
We conclude that the zero does not disappear and follows a smooth path parameterized by $t$.

An adiabatic evolution, by definition, must be invertible. If a zero appeared at some $t=T$, then in the inverse process, the zero would disappear. Since this process is also continuous, this would contradict our previous statement. Therefore, we arrive at the conclusion that the multiplicity of zeros of $\alpha_b^{(m)}(z)$ cannot be changed adiabatically for $z$ nonzero and finite. In other words, in an infinite system, Bloch states satisfying \eqref{eq:edge state boundary condition} cannot be created or destroyed in this range: they can only shift around in $z$ space.

This justifies our focus on Bloch states with $\alpha_b^{(m)}(z)=0$: although those states might seem improbable, if some system has such a Bloch state, every system to which it is adiabatically connected also has a Bloch state with that sublattice zero at some value of $z$.

\begin{figure}[t]
    \centering
    \includegraphics[width=0.75\linewidth]{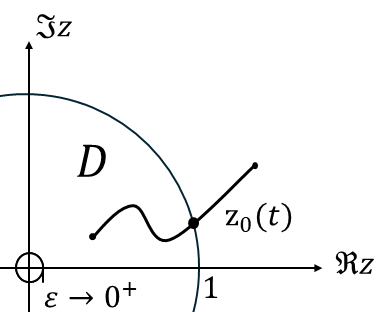}
    \caption{A possible trajectory, in complex $z$ space, of a simple zero of $\alpha_b^{(m)}(z)$ during an adiabatic evolution. The large dot represents the point where $|z_0(t)|=1$, at which the non-analytically continued Bloch Hamiltonian has a state with a sublattice zero. In other words, at this point, the edge state becomes a bulk state: it crosses the band. In the scenario shown above, the edge state would disappear from the semi-infinite chain after crossing the $|z|=1$ circle.}
    \label{fig:z0(t)}
\end{figure}

\subsection{From infinite to semi-infinite}
\label{subsec:InfiniteToSemi}

We have just argued that the zeros of the function $\alpha_b^{(m)}(z)$ cannot be created or destroyed for $z$ nonzero and finite. We know that if $\alpha_b^{(m)}(z)=0$, then a boundary condition of the form \eqref{eq:edge state boundary condition} is satisfied. However, for that state to be normalizable, it must have $|z|\leq 1$. Nothing prohibits the zeros of $\alpha_b^{(m)}(z)$ from crossing the unit circle $|z|=1$ adiabatically. Therefore, generally, edge states respecting the boundary condition \eqref{eq:edge state boundary condition} are not adiabatically protected, as is the case in the SSH3 model~\cite{SSH3}. A possible path of a zero during an adiabatic evolution, $z_0(t)$, is shown in \autoref{fig:z0(t)}.

Before continuing, we must address the fact that, $\alpha_b^{(m)}(z)$ being non-analytic at $z=0$ and $\infty$, it seems like new edge states can appear and disappear at these points. In our  system, a state with $z=0$ is one entirely localized in the first unit cell. An acceptable state of the system can only arise at such a value of $z$ when there are no inter-cell hoppings (otherwise \eqref{eq:Analytic continuations} will have an ``acceptable'' state with infinite energy). New edge states can appear and disappear when there are no inter-cell couplings (what might be called the ``$N$-merized'' limit); for this reason we shall restrict the adiabatic evolutions that we consider to those that avoid this limit.

It is important to understand how adiabatic protection fits in our picture. If the $z$ of an edge state crosses the unit circle, $|z|=1$, it will go from normalizable to non-normalizable and thus disappear from the semi-infinite system. For adiabatically protected states, we require that a gap close when this happens. This is true if the edge state belongs to multiple bands at once or, in other words, if $\ket{\alpha^{(m)}(z_{e.p.})}=\ket{\alpha^{(n)}(z_{e.p.})}$ with $m\neq n$ and some $z_{e.p.} \in \mathbb{C}$. This can happen at exceptional points of the analytically continued Bloch Hamiltonian. Those exceptional points correspond to two (or more) states coalescing at $z_{e.p.}$, making $h(z)$ non-diagonalizable at that point.\footnote{In the literature there exists a number of definitions of the term \emph{exceptional point}. We are only concerned with this specific type.} We will refer to the states that coalesce at an exceptional point as \emph{exceptional states}. When such a state crosses $|z|=1$, there must be a band crossing, since this state has a real $k$, meaning that for some $m\neq n$, $E^{(m)}(k_{e.p.})=E^{(n)}(k_{e.p.})$.

\subsection{Exceptional points}
\label{subsec:exceptionalPoints}

Through what has been said above, we can see that the edge states that satisfy the boundary condition \eqref{eq:edge state boundary condition} and arise at exceptional points are adiabatically protected (provided the edge state remains exceptional throughout the adiabatic evolution; this will be addressed in \autoref{sec:chiralSymmetry}). It is useful to understand how to count the number of exceptional points.

Consider equation \eqref{eq:scalar product}. When two different states coalesce at an exceptional point,
\begin{align}
    0&=\bra{\alpha^{(m)}(1/z_{e.p.}^*)}\alpha^{(n)}(z_{e.p.})\rangle\nonumber\\
    &\propto \bra{\alpha^{(m)}(1/z_{e.p.}^*)}\alpha^{(m)}(z_{e.p.})\rangle
    \label{eq:self-orth}
\end{align}
where the first line (with $m\neq n$) follows directly from \eqref{eq:scalar product} and the second follows from the fact that, at an exceptional point, $|\alpha^{(n)}(z_{e.p.})\rangle \propto |\alpha^{(m)}(z_{e.p.})\rangle$. This phenomenon is called self-orthogonality~\cite{Nimrod}. Note that we assume a normalization such that $|\alpha^{(n)}(z)\rangle$ does not have poles; otherwise, we could choose
\[
|\hat{\alpha}^{(n)}(z)\rangle \equiv \Big(\bra{\alpha^{(n)}(1/z^*)}\alpha^{(n)}(z)\rangle\Big)^{-1/2}|\alpha^{(n)}(z)\rangle,
\]
which will not satisfy \eqref{eq:self-orth}.

The para-Hermitian norm
\begin{equation}
    \mathcal{N}^{(m)}=\bra{\alpha^{(m)}(1/z^*)}\alpha^{(m)}(z)\rangle
\end{equation}
is analytic up to branch cuts~\cite{Topology_of_exceptional_points, Physics_of_E.P., Kato}. We can use this property to count its zeros in $D$. Each of them will correspond to a self-orthogonal exceptional point. To do so, we will again use the argument principle:
\begin{equation}\label{eq:winding of N}
    \gamma^{(m)} \equiv -i \oint_{\partial D} dz \, \partial_z \ln(\mathcal{N}^{(m)}(z)),
\end{equation}
{\color{red}%
where, for exceptional points of order $r=2$ (the generic case, and the only one we consider from now on), $\gamma^{(m)}/2\pi$ is half the number of exceptional points at which band $m$ participates, since $\mathcal{N}^{(m)}(z)$ vanishes as $(z-z_{e.p.})^{1/2}$ there.\footnote{As has been mentioned previously, if $r$ states coalesce, the branch cut will be caused by a root of order $r$ \cite{Higher_order_E.P.}.} Since two bands coalesce at each such point, summing over bands counts each exceptional point once, and the total number of exceptional points is
}%end red
\begin{equation}
    \text{\# e.p.}=\frac{1}{2\pi}\sum_{m=1}^{N} \gamma^{(m)}.
\end{equation}

Interestingly, each $\gamma^{(m)}$ can be written in terms of the analytically continued Berry connection
\begin{equation}
    \mathcal{A}^{(m)}(z) = i\frac{ \bra{\alpha^{(m)}(1/z^*)}\partial_z \alpha^{(m)}(z)\rangle}{{\bra{\alpha^{(m)}(1/z^*)}\alpha^{(m)}(z)\rangle}}.
\end{equation}
Let $S(R)$ be a circle of radius $R$ centred at $z=0$. Then
\begin{align}
\gamma^{(m)}=\lim_{\varepsilon\to 0^+}\Bigg[&\left(\oint_{S(\varepsilon)}dz \mathcal{A}^{(m)}(z)\right)^*\nonumber\\
&\quad-\oint_{S(\varepsilon^{-1})}dz \mathcal{A}^{(m)}(z)\Bigg].
\label{eq:Zak phase}
\end{align}

\subsubsection{Connection to topology}
\label{subsubsec:topology}
We can organize the eigenstates of $h(z)$ into a matrix and count the number of exceptional points in this way. We define
\begin{equation}\label{eq:U}
    U(z)=
    \begin{pmatrix}
        \renewcommand{\arraystretch}{1.3}
        \begin{array}{@{}c|c|c|c@{}}
            \ket{\alpha^{(1)}(z)} & \ket{\alpha^{(2)}(z)} & \cdots & \ket{\alpha^{(N)}(z)}
        \end{array}
    \end{pmatrix}.
\end{equation}
Since we have chosen a gauge and normalization such that $\ket{\alpha^{(m)}(z)}$ is analytic (up to branch cuts) and nonzero, $U(z)$ is itself analytic up to branch cuts. Furthermore, $\det{U(z)}=0$ only when two states coalesce, thus
\begin{align}
        \text{\# e.p.}&=\frac{r}{2\pi i}\oint_{\partial D}dz \, \partial_z \ln (\det U(z))\nonumber\\
        &=\frac{r}{2\pi i}\oint_{\partial D}dz \Tr \Big(U^{-1}(z)\partial_z U(z)\Big),
        \label{eq:trU}
\end{align}
where in the second line we have used Jacobi's formula. Here, a factor $r$ appears in the numerator because each exceptional point will contribute $1/r$ to this integral. Note that we assume $U(z)$ to be invertible on $\partial D$. The matrix $U(z)$ is invertible when $h(z)$ is diagonalizable, so we will also assume that $h(z)$ is diagonalizable on the integration path.

Evaluated along a loop, $U(z)$ is a mapping $S^1 \to GL(N,\mathbb{C})$. This allows us to interpret \eqref{eq:trU} in terms of homotopy between $S^1$ and $GL(N,\mathbb{C})$.

{\color{red}%
Generically, around an exceptional point of order $2$, the coalescing eigenvectors of $h(z)$ admit the (Puiseux) expansion~\cite{Guenther2007, Physics_of_E.P.}
\begin{equation}\label{eq:Puiseux}
    |\alpha_{\pm} (z)\rangle \sim |\alpha_0 \rangle \pm \sqrt{z-z_{e.p.}}|\alpha_1 \rangle  + ...\, ,
\end{equation}
therefore, as the state $|\alpha_{\pm}\rangle$ is transported through a loop encircling the exceptional point, $z-z_{e.p.} = \rho e^{i\theta}$ with $\theta\in [0, 2\pi)$, it does not go back to its original value, and $U(z)$ has a non-trivial monodromy. We will limit our analysis to $r=2$ which is the relevant case for the following sections. When $\theta$ goes from $0$ to $2\pi$, $|\alpha_{\pm}(z)\rangle$ switches its branch, and $\theta=2\pi \to 4\pi$ sends it back to the original branch. The evaluation of the homotopy of $U(z)$ requires $U(z)$ to be single valued, continuous and periodic on the contour. To do so, we define $U(z)$ on two copies of $\mathbb{C}$ glued together along the branch cut, and evaluate the winding of $U(z)$ on two copies of $S(R)$ (connected to each other if they pass through a branch cut), one on each Riemann sheet. This contour will be referred to as $2S(R)$.

Taking monodromy into consideration, we define the winding number in the presence of an exceptional point as
\begin{equation}\label{eq:w(R)}
    w(R)\equiv \frac{1}{2\pi i}\oint_{2S(R)}dz \Tr \Big(U^{-1}(z)\partial_z U(z)\Big).
\end{equation}
}%end red
This winding represents a homotopy class of $\pi_1(GL(N,\mathbb{C})) \cong \mathbb{Z}$~\cite{Topology_of_exceptional_points}.

\textcolor{red}{When switching Riemann sheets, $|\alpha_{\pm}\rangle \to -|\alpha_{\mp}\rangle $, that is to say, two columns of $U(z)$ are interchanged and multiplied by $-1$. This changes the sign of $\det U(z)$, but it has no effect on \eqref{eq:w(R)}. Therefore we can substitute $2S(R)\to S(R)$ and $\frac{1}{2\pi i}\to \frac{1}{\pi i}$ in (\ref{eq:w(R)}).}

In terms of $w(R)$, we can rewrite \eqref{eq:trU} as
\begin{equation}
    \text{\# e.p.} =  \lim_{\varepsilon\to 0^+}(w(1)-w(\varepsilon)).
\end{equation}
If $\text{\#e.p.} \neq 0$, it means that there is a discontinuity in the mappings $U(S(R))$, since two mappings with different $R$ have a different homotopy. This discontinuity occurs because, at the exceptional point, $U$ is no longer invertible and the mapping is not defined. \textcolor{red}{The difference of windings $w(R)-w(0^+)$ is gauge invariant since it simply counts the number of zeros of $\det U(z)$, which cannot be changed by a choice of gauge.}

Although our initial investigation of edge states relied on the validity of the boundary condition \eqref{eq:edge state boundary condition}, allowing us to emphasize the importance of zeros of the wave function, we have now gained the insight that the number of exceptional points of an analytically continued Bloch Hamiltonian is adiabatically protected. Therefore, if one is given a boundary condition (not necessarily \eqref{eq:edge state boundary condition}) and wishes to construct an adiabatically protected state, one can do so by combining exceptional states. The number of exceptional points is thus directly related to the number of edge states.

\subsection{Chiral symmetry}
\label{sec:chiralSymmetry}

Although we have established that edge states arising at exceptional points are adiabatically protected, we have not yet explained under which circumstances exceptional points do appear. One such circumstance arises if the Hamiltonian has chiral symmetry, that is, if a constant, unitary, Hermitian matrix $\Gamma$ exists which anticommutes with the $N$-dimensional Bloch Hamiltonian: $\{\Gamma,h(z)\} = 0$. With a change of basis, $h(z)$ and $\Gamma$ can be brought to the form
\begin{align}
\label{eq:chiral hamiltonian}
h(z) &=
\begin{pmatrix}
    % some tricks to get columns of equal width (\matrixcolwidth defined in preamble)
    \renewcommand{\arraystretch}{1.3}
    \settowidth{\matrixcolwidth}{$Q(1/z^*)^\dagger$}%use widest matrix element
    \begin{array}{w{c}{\matrixcolwidth} | w{c}{\matrixcolwidth}}
        0 & Q(z) \\
        \hline
        Q(1/z^*)^\dagger & 0
    \end{array}
\end{pmatrix},
\\
\label{eq:Gamma}
\Gamma &=
\begin{pmatrix}
    % some tricks to get columns of equal width (\matrixcolwidth defined in preamble)
    \renewcommand{\arraystretch}{1.3}
    \settowidth{\matrixcolwidth}{$-1$}%use widest matrix element
    \begin{array}{w{c}{\matrixcolwidth} | w{c}{\matrixcolwidth}}
        \mathds{1} & 0 \\
        \hline
        0 & -\mathds{1}
    \end{array}
\end{pmatrix}.
\end{align}
Chiral symmetry divides the system into sublattices of positive and negative chirality, which we will refer to as sublattice $A$ and $B$, respectively. A priori, the two sublattices need not contain the same number of sites per unit cell: if they contain $N_A$ and $N_B$ sites ($N_A+N_B=N$), respectively, then $Q(z)$ is an $N_A\times N_B$ matrix, and the identity blocks in $\Gamma$ have dimensions $N_A$ and $N_B$.
The energy spectrum is symmetric; in the above basis, eigenstates of equal and opposite nonzero energy are given by
\begin{equation}
\ket{\chi_\pm(z)} =
\begin{pmatrix}
    \chi_{A}(z)\\
    \pm\chi_{B}(z)
\end{pmatrix}
\end{equation}
with $\chi_A$ and $\chi_B$ nonzero.

If either $\chi_A$ or $\chi_B$ is zero, the two states are identical (up to a sign), and the state must have zero energy. If $N_A\neq N_B$ (in particular, if $N$ is odd), this is guaranteed to occur for all $z$: $Q(z)$ is then not square and $h(z)$ has at least $|N_A-N_B|$ zero-energy flat bands. Since such zero-modes are unrelated to the considerations in this paper, we will assume henceforth that $N_A=N_B=N/2$ with $N$ even, in which case all submatrices in \eqref{eq:chiral hamiltonian} and \eqref{eq:Gamma} are square matrices of dimension $N/2$.

With $N_A=N_B$, while zero-modes are not required, it is entirely possible that at isolated values of $z$ either $\chi_A$ or $\chi_B$ is zero, again giving a zero-mode. Such values of $z$ and the associated zero-modes are the exceptional points and exceptional states referred to in the previous subsections.

{\color{red}%
We assume that the zeros of $\det Q(z)$ and of $\det Q(1/z^*)^\dagger$ are simple and do not coincide, which is the generic situation; at a common zero, both blocks are singular and $h(z)$ has a two-dimensional kernel without being defective, while repeated zeros can give rise to higher-order exceptional points.

Under these generic assumptions, $r=2$: the coalescing states come in pairs.
}%end red
These exceptional states always have half of their components fixed to zero; therefore they always satisfy a boundary condition of the form \eqref{eq:edge state boundary condition}. However, as has been discussed, the number of exceptional states can give us the number of adiabatically protected edge states for boundary conditions of other forms as well.

Equation \eqref{eq:chiral hamiltonian} implies that
\begin{align}
\label{eq:dog3}
    E^{(m)}(z) \chi_{A}^{(m)}(z) &= Q(z)\chi_{B}^{(m)}(z)\\
    E^{(m)}(z) \chi_{B}^{(m)}(z) &= Q(1/z^*)^\dagger\chi_{A}^{(m)}(z).
\end{align}
If $\chi^{(m)}_{A/B}(z_0)=0$, then $E^{(m)}(z_0)=0$. The boundary of a semi-infinite system has to fall either in subspace $A$ or in subspace $B$. We know that when $\chi_{A/B}=0$, $h(z)$ has an exceptional point. When this happens, we also satisfy a boundary condition; therefore, chiral symmetry protected zero energy edge states arise at the exceptional points of $h(z)$, at least when the hoppings going through the boundary are to the nearest neighbour.

The matrix $U(z)$ from \eqref{eq:U} is given by
\begin{align}
    U(z)&=
    \begin{pmatrix}
        \renewcommand{\arraystretch}{1.3}
        \begin{array}{@{}c|c|c|c|c|c@{}}
            \chi_A^{(1)} & \cdots & \chi_A^{(N/2)} & \chi_A^{(1)} & \cdots & \chi_A^{(N/2)}\\
            \hline
            \chi_B^{(1)} & \cdots & \chi_B^{(N/2)} & -\chi_B^{(1)} & \cdots & -\chi_B^{(N/2)}
        \end{array}
    \end{pmatrix}\nonumber\\
    &\equiv
    \begin{pmatrix}
    % some tricks to get columns of equal width (\matrixcolwidth defined in preamble)
        \renewcommand{\arraystretch}{1.3}
        \settowidth{\matrixcolwidth}{$-U_B(z)$}%use widest matrix element
        \begin{array}{w{c}{\matrixcolwidth} | w{c}{\matrixcolwidth}}
            U_A(z) & U_A(z)\\
            \hline
            U_B(z) & -U_B(z)
        \end{array}
    \end{pmatrix}.
    \label{eq:chiral U}
\end{align}
The block structure of $U(z)$ allows us to separate the exceptional points with $\chi_A=0$ from those with $\chi_B=0$. Indeed, adding the second block-column of \eqref{eq:chiral U} to the first gives
\begin{equation}
    \det U(z) = (-2)^{N/2}\det U_A(z)\,\det U_B(z).
\end{equation}
Since $\det U(z)\neq0$ wherever $h(z)$ is diagonalizable, $\det U_A(z)$ and $\det U_B(z)$ can only vanish at exceptional points. Moreover, the columns of $U_{A/B}(z)$ are the components $\chi_{A/B}^{(n)}(z)$ of the eigenstates, so that at an exceptional point with $\chi_A=0$ ($\chi_B=0$), a column of $U_A(z)$ ($U_B(z)$) vanishes, as does $\det U_A(z)$ ($\det U_B(z)$).

Note that $U(z)$ is analytic up to branch cuts; we can therefore count the exceptional states with $\chi_{A/B} = 0$ using the argument principle,
\begin{align}
    \nu_{A/B} &= \frac{1}{2 \pi i}\oint_{\partial D}dz \, \textcolor{red}{\partial_z} \ln \det U_{A/B}(z)\nonumber\\
    &=\frac{1}{2 \pi i}\oint_{\partial D}dz \Tr \Big( U_{A/B}^{-1}(z)\partial_z U_{A/B}(z)\Big).
    \label{eq:cool equation}
\end{align}
As before, this expression can be understood in terms of the $GL(N/2,\mathbb{C})$ homotopy classes.
{\color{red}%
Due to the square root, $2\nu_{A/B}$ is the number of exceptional states with $\chi_{A/B}=0$ in $D$.

Some remarks on \eqref{eq:cool equation} are in order. The domain $D$ is the punctured disk $0<|z|<1$, whose boundary consists of the unit circle, traversed anticlockwise, and a circle of radius $\varepsilon\to0^+$ around the origin, traversed clockwise. On the unit circle, $h(z)$ is Hermitian and, the system being gapped, its eigenvalues are real and nondegenerate, so that the bands can be labelled by their energies and $U_{A/B}(z)$ is single-valued there. On the small circle around the origin, as around any exceptional point, the columns of $U_{A/B}(z)$ may be permuted (up to signs); as in \autoref{subsubsec:topology}, this changes $\det U_{A/B}(z)$ at most by a sign, so that the integrand $\partial_z\ln\det U_{A/B}(z)$ is single-valued on both contours. We choose a normalization and gauge in which the eigenstates are analytic and nonvanishing in $D$ except at branch points, so that the zeros of $\det U_{A/B}(z)$ in $D$ are precisely the exceptional points with $\chi_{A/B}=0$. Under the genericity assumptions stated above, each of these is a simple zero of $\det Q(z)$ (for $\chi_A=0$) or of $\det Q(1/z^*)^\dagger$ (for $\chi_B=0$), near which $\det U_{A/B}(z)$ vanishes as a square root and contributes $1/2$ to \eqref{eq:cool equation}. The origin is excluded from $D$, so that the poles of $h(z)$ at $z=0$, whose order depends on the choice of unit cell, do not contribute. Finally, the number of zeros in $D$ can change only if a zero crosses the unit circle, which requires the gap to close, or reaches $z=0$ or $z=\infty$; the latter happens only if the coefficient of the highest or lowest power of $z$ in $\det Q(z)$ vanishes (as for $t_3\to0$ in \autoref{sec:extendedSSH}), and we exclude such deformations.
}%end red

To compute \eqref{eq:cool equation}, we need to fix a gauge and a normalization that makes $\ket{\alpha^{(m)}(z)}$ analytic up to branch cuts. Often, we do not have control over the gauge of the eigenstates when performing numerical calculations. A more computable quantity is
\begin{equation}
    \nu=\nu_{B}-\nu_A.
\end{equation}

The winding number $\nu$ does not require an analytic choice of gauge, nor does it require $\ket{\alpha^{(m)}(z)}$ to have a normalization that avoids poles, since any choice of normalization and gauge of the eigenstates affects $U_A$ and $U_B$ in the same way, and therefore cancels in $\nu$.
From \eqref{eq:dog3}, $Q(z)=U_A(z)\Sigma(z)U_B^{-1}(z)$, where $\Sigma(z)=\mathrm{diag}\bigl(E^{(1)}(z),\dots,E^{(N/2)}(z)\bigr)$~\cite{Chiral_winding}.
This expression is unchanged if each eigenstate is normalized, so we may replace $U_{A/B}(z)$ by the matrices $\hat{U}_{A/B}(z)$ defined by

\begin{multline}
    \begin{pmatrix}
        \renewcommand{\arraystretch}{1.3}
        \begin{array}{@{}c|c|c|c@{}}
            \ket{\hat\alpha^{(1)}(z)} & \ket{\hat\alpha^{(2)}(z)} & \cdots & \ket{\hat\alpha^{(N)}(z)}
        \end{array}
    \end{pmatrix}\\
    =
    \begin{pmatrix}
% some tricks to get columns of equal width (\matrixcolwidth defined in preamble)
        \renewcommand{\arraystretch}{1.3}
        \settowidth{\matrixcolwidth}{$-\hat{U}_B(z)$}%use widest matrix element
        \begin{array}{w{c}{\matrixcolwidth} | w{c}{\matrixcolwidth}}
            \hat{U}_A(z) & \hat{U}_A(z)\\
            \hline
            \hat{U}_B(z) & -\hat{U}_B(z)
        \end{array}
    \end{pmatrix},
\end{multline}
where $\ket{\hat\alpha^{(n)}(z)}$ are the para-normalized states defined in \autoref{subsec:exceptionalPoints}.

By defining $q(z)=\hat{U}_A(z)\hat{U}_B^{-1}(z)$, we can write
\begin{equation}\label{eq:chiral winding}
    \nu=-\frac{1}{2\pi i }\oint_{\partial D} dz \Tr \Big( q^{-1}(z) \partial_z q(z)\Big),
\end{equation}
which is the usual chiral winding number~\cite{tr_Q}, although integrated on a slightly different domain (usually the integral is performed on $S(1)$). Note that $q(z)$ is $Q(z)$ in the flat band limit.

{\color{red}%
The winding number $\nu$ is the signed topological index: it gives the difference between the numbers of exceptional states with $\chi_B=0$ and with $\chi_A=0$ in $D$, which is an adiabatic invariant, but not the number of edge states for a given termination. The latter is determined by $2\nu_A$ and $2\nu_B$ separately, together with the boundary conditions, as discussed in \autoref{subsec:movingAway}.
}%end red
This is consistent with the well-known fact that the chiral winding number predicts the number of edge states only for a suitable choice of unit cell.

For the same reasons as discussed at the end of \autoref{subsec:edgeStatesWithoutBranchCuts}, the winding numbers presented here do not depend on the choice of unit cell.

We have, in this section, shown the existence of chiral symmetry protected zero energy states in the Bloch Hamiltonian. Those states can be used to construct edge states of a semi-infinite system. In particular, if the boundary condition \eqref{eq:edge state boundary condition} is sufficient, \eqref{eq:cool equation} directly counts the number of edge states of the semi-infinite chain.

\subsection{Moving away from nearest-neighbour hoppings across the boundary}
\label{subsec:movingAway}
As discussed in \autoref{sec:bulkBoundaryStates}, the boundary condition
\begin{equation}\label{eq:boundary condition}
    \langle p_b,\mu_b | \Psi \rangle = 0
\end{equation}
is only sufficient if the only hoppings crossing the boundary (in the corresponding infinite system) are to the nearest neighbour. In general, for hopping range $R$ there will be up to $R$ boundary conditions, and \eqref{eq:boundary condition} is not necessarily sufficient. We will now generalize the treatment above to such systems.

By studying the zeros of the Bloch wavefunction $|\alpha(z)\rangle$ we have gained the important insight that chiral symmetry leads to adiabatically protected states with $\chi_A=0$ or $\chi_B=0$, due to the presence of exceptional points. When the boundary condition \eqref{eq:edge state boundary condition} is sufficient, it so happens that such exceptional states are also edge states. Counting the number of Bloch states with $\chi_A=0$ or $\chi_B=0$ is not necessarily sufficient to count the number of states fully satisfying the boundary condition, but it allows one to count the dimension of the set of candidate states which can be used to construct an adiabatically protected edge state. Additionally, through the $\chi_A=0$ and $\chi_B=0$ conditions, the exceptional states of chiral symmetric Hamiltonians start with at least half of the boundary conditions already being satisfied. Knowing the number of remaining boundary conditions and the dimension of the $\chi_A=0$ and $\chi_B=0$ subspaces, one can predict some edge states of a semi-infinite system. \textcolor{red}{If one could show that all edge states must arise at an exceptional point, this would give the full number of edge states. Without this knowledge, we can only use the number of exceptional points and boundary conditions to give a lower bound on the number of adiabatically protected edge states.}

In general, one should interpret the windings $\nu_{A/B}$ as counting the dimensions of the spaces of adiabatically protected Bloch states which can be used to construct a semi-infinite topological edge state.
\textcolor{red}{%
If there are $N_\text{b}'$ boundary conditions left to satisfy in the $\chi_A=0$ or $\chi_B=0$ subspaces, the number of edge states from either subspace is bounded from below:
\begin{equation}
\label{eq:edge-state-count}
    \text{\# e.s.} \geq 
    \begin{cases}
        0 &\text{if }2\nu_{A/B} \leq N_\text{b}',\\
        2\nu_{A/B} - N_\text{b}' &\text{if }2\nu_{A/B} > N_\text{b}'.
    \end{cases}
\end{equation}
}%end red
Whenever the dimension of the space of available exceptional points changes, the gap closes and the number of edge states changes. A more detailed analysis of the possible boundary conditions is presented in App.~\ref{app:boundarySemiInfinite}.
These results are illustrated on an extended SSH model in the next section
%formerly \autoref{sec:extendedSSH}, so beware if add eg a new section before the next one
and on a four-band model in App.~\ref{app:4band}.

\section{Example: extended SSH model}
\label{sec:extendedSSH}

We will illustrate our construction through a generalization of the SSH model that has already been studied in~\cite{Zeros_of_order_2}. Consider an extension of the SSH model shown in \autoref{fig:extended_SSH}, with real hopping parameters $t_1,t_2,t_3$ and with range $R=3$. For $t_3=0$, it reduces to the standard SSH model, with two sites per unit cell and alternating nearest-neighbour hoppings $t_1$ and $t_2$. This model has analytically continued Bloch Hamiltonian
\begin{equation}
    h(z)=
    \begin{pmatrix}
        0 & t_1+t_2 z^{-1}+t_3 z^{-2}\\
        t_1+t_2 z + t_3z^2 & 0
    \end{pmatrix},
\end{equation}
which is non-diagonalizable when $t_1+t_2 z_{e.p.}^{-1}+t_3 z_{e.p.}^{-2}=0$ or $t_1+t_2 z_{e.p.}^{1}+t_3 z_{e.p.}^{2}=0$. Its eigenstates, in a gauge and normalization where they are analytic up to branch cuts
and non-vanishing, are given by
\begin{equation}\label{eq:eigenstates}
    \ket{\pm(z)}\to
    \begin{pmatrix}
        \sqrt{t_1+t_2 z^{-1}+t_3 z^{-2}}\\
        \pm\sqrt{t_1+t_2 z^{1}+t_3 z^{2}}
    \end{pmatrix}
\end{equation}
and its energies are 
\begin{equation}
    E_{\pm}(z)=\pm\sqrt{(t_1+t_2 z^{-1}+t_3 z^{-2})(t_1+t_2 z^{1}+t_3 z^{2})}.
\end{equation}
In agreement with \eqref{eq:EnotReal} and \eqref{eq:scalar product}, $E_{\pm}(z)=(E_{\pm}(1/z^*))^*$ and $\bra{\pm(1/z^*)}\mp(z)\rangle=0$. Furthermore, the two eigenstates coalesce at the non-diagonalizable points of $h(z)$ at which $\bra{+(1/z_{e.p.}^*)}+(z_{e.p.})\rangle=\bra{-(1/z_{e.p.}^*)}-(z_{e.p.})\rangle=0$.

\begin{figure}[t]
  \centering
  \resizebox{\linewidth}{!}{%
  \begin{tikzpicture}[
      site/.style={circle,fill=black,inner sep=1.4pt},
      lab/.style={font=\footnotesize},
      tone/.style={line width=1.3pt},
      ttwo/.style={line width=0.5pt},
      tthree/.style={line width=0.5pt,dash pattern=on 3pt off 2pt}]
    \def\dx{0.85}
    % sites: A_p at even k, B_p at odd k
    \foreach \k in {0,...,9}{
      \pgfmathtruncatemacro{\m}{mod(\k,2)}
      \node[site] (s\k) at (\k*\dx,0) {};
      \ifnum\m=0 \def\lbl{$A$}\else \def\lbl{$B$}\fi
      % endpoints of the lower t3 arcs (k=3,6) are labelled above the chain
      \ifnum\k=3
        \node[lab,above=1pt] at (s\k) {\lbl};
      \else\ifnum\k=6
        \node[lab,above=1pt] at (s\k) {\lbl};
      \else
        \node[lab,below=1pt] at (s\k) {\lbl};
      \fi\fi}
    % t1: A_p - B_p (within cell); t2: B_p - A_{p+1} (between cells)
    \foreach \k in {0,2,...,8}{
      \pgfmathtruncatemacro{\kb}{\k+1}
      \draw[tone] (s\k) -- (s\kb);}
    \foreach \k in {1,3,...,7}{
      \pgfmathtruncatemacro{\kb}{\k+1}
      \draw[ttwo] (s\k) -- (s\kb);}
    % t3: B_p - A_{p+2} (range 3), alternately above and below
    \draw[tthree] (s1) to[out=75,in=105,looseness=1.25] (s4);
    \draw[tthree] (s3) to[out=-75,in=-105,looseness=1.25] (s6);
    \draw[tthree] (s5) to[out=75,in=105,looseness=1.25] (s8);
    % unit cell
    \draw[gray,rounded corners=3pt,line width=0.4pt]
      (4*\dx-0.3,0.5) rectangle (5*\dx+0.3,-0.5);
    % legend
    \begin{scope}[xshift=8.6cm,yshift=0.5cm]
      \draw[tone]   (0,0)    -- (0.6,0);    \node[right] at (0.6,0)    {$t_1$};
      \draw[ttwo]   (0,-0.5) -- (0.6,-0.5); \node[right] at (0.6,-0.5) {$t_2$};
      \draw[tthree] (0,-1.0) -- (0.6,-1.0); \node[right] at (0.6,-1.0) {$t_3$};
    \end{scope}
  \end{tikzpicture}}
  \caption{An extension of the SSH model, with hoppings $t_1$ within a unit cell (thick line) and $t_2$, $t_3$ between unit cells (thin solid and dashed lines). The box indicates a unit cell.}
  \label{fig:extended_SSH}
\end{figure}
Consider the corresponding semi-infinite chain. There are two possible boundaries for a chain with a boundary on the left:
\begin{enumerate}
    \item The first site is an element of sublattice $A$.
    \item The first site is an element of sublattice $B$.
\end{enumerate}
According to the discussion of App.~\ref{app:boundarySemiInfinite}, if the first site of the system is an $A$-site, the boundary condition is that the first and third sites outside the chain must be zero, a condition which is naturally fulfilled if $B=0$. If the first site is a $B$-site, the first and second sites outside the chain must be zero. If we want to build an adiabatically protected state that fulfills this condition, we need either two states with $B=0$, or two states with $A=0$ in $D$. The amplitude of the $B$-site can be obtained from that of the $A$ site by replacing $z$ with $1/z$; if $A$ has two zeros in $D$, $B$ has none and vice versa. Since the amplitude of $A$ is the square root of a quadratic polynomial, boundary 1 admits between 0 and 2 adiabatically protected edge states, whereas boundary 2 only admits between 0 and 1.

We can compute the number of exceptional states with either $A=0$ or $B=0$ using \eqref{eq:SZP}. Alternatively, we can write down \eqref{eq:U} and then compute \eqref{eq:cool equation}. Here,
\begin{equation}\label{eq:U(z) exemple}
    U(z)=
    \begin{pmatrix}
        \sqrt{t_1+t_2 z^{-1}+t_3 z^{-2}} & \sqrt{t_1+t_2 z^{-1}+t_3 z^{-2}}\\
        \sqrt{t_1+t_2 z^{1}+t_3 z^{2}} & -\sqrt{t_1+t_2 z^{1}+t_3 z^{2}}
    \end{pmatrix},
\end{equation}
from which we see that 
\begin{equation}
    U_A(z)=\sqrt{t_1+t_2 z^{-1}+t_3 z^{-2}}
\end{equation}
and
\begin{equation}
    U_B(z)=\sqrt{t_1+t_2 z^{1}+t_3 z^{2}}.
\end{equation}
Here, $U_A$ and $U_B$ simply give the components of the wavefunction; therefore, the two approaches are exactly the same. The usual Zak phase is given by the difference of $U_A$ and $U_B$ windings.

\textcolor{red}{%
In this example, we can explicitly see that, as $z$ goes around an exceptional point, the columns of $U(z)$ get interchanged. The term $\sqrt{t_1+t_2z^{-1}+t_3z^{-2}}$ can be written $z^{-1}\sqrt{t_1 z^2+t_2 z + t_3}$. Let $z_0$ be a zero of $t_1 z^2+t_2 z + t_3$. By switching to the coordinates $z=z_0+\varepsilon$ and expanding in $\varepsilon$, we obtain
\begin{align}
    &\sqrt{t_1+t_2z^{-1}+t_3z^{-2}}\\
    &= \frac{1}{z_0+\varepsilon}\varepsilon^{1/2}\sqrt{t_2+2t_1z_0}\sqrt{1+\frac{t_1}{t_2+2t_1z_0}\varepsilon}\nonumber\\
    &=\frac{\sqrt{t_2+2t_1z_0}}{z_0}\varepsilon^{1/2}(1+O(\varepsilon))\nonumber.
\end{align}
Therefore, going a full circle around an exceptional point leads to $\sqrt{t_1+t_2z^{-1}+t_3z^{-2}} \to -\sqrt{t_1+t_2z^{-1}+t_3z^{-2}}$ and leaves $\sqrt{t_1+t_2 z^{1}+t_3 z^{2}}$ unchanged. The net effect is to switch the two columns of $U(z)$ and multiply it by $-1$. The exact same argument, with the same conclusion, can be repeated for exceptional points that arise at the zeros of $\sqrt{t_1+t_2 z^{1}+t_3 z^{2}}$. The column interchange changes the sign of $\det U(z)$ (the overall factor $-1$ does not, since $U$ is a $2\times 2$ matrix), but this drops out of $\partial_z \ln\det U(z)$, so this peculiarity has no effect on the computation of winding numbers.
}%end red

The number of exceptional states with $A=0$ or $B=0$ in $D$ is given respectively by
\begin{equation}
    2\nu_{A} = \frac{1}{\pi i} \oint_{\partial D} dz \, \partial_z \ln (\sqrt{t_1+t_2 z^{-1}+t_3 z^{-2}})
\end{equation}
and
\begin{equation}
    2\nu_{B} = \frac{1}{\pi i} \oint_{\partial D} dz \, \partial_z \ln (\sqrt{t_1+t_2 z^{1}+t_3 z^{2}}).
\end{equation}
Each varies between 0 and 2 depending on the parameters $t_1,t_2$ and $t_3$. Their nature also varies from purely exponential ($z \in \mathbb{R}$)
to oscillating exponential ($z \notin \mathbb{R}$). The phase diagram describing the number of exceptional points with $B=0$ is shown in \autoref{fig:phase diagram}.

\begin{figure}[htb]
  \centering
  \begin{tikzpicture}[x=1cm,y=1cm,
      bdry/.style={line width=0.5pt},
      % region styles (same order as the legend)
      rW/.style={fill=white},                                  % white
      rG/.style={fill=black!45},                               % grey
      rB/.style={pattern={Lines[angle=45,distance=3.5pt,line width=0.4pt]},
                 pattern color=black!70},                      % blue
      rR/.style={fill=black!25},                               % red strip
      rGr/.style={pattern={Lines[angle=-45,distance=3.5pt,line width=0.4pt]},
                  pattern color=black!70},                     % green
      % lines on the axes
      axIn/.style={line width=1.8pt},                          % red line
      axOut/.style={line width=2.2pt,line cap=round,dash pattern=on 0pt off 4.5pt}, % orange line (dots)
      halo/.style={white,line width=4pt}]
    \def\xmin{-4} \def\xmax{4.2} \def\ymin{-3.8} \def\ymax{3.7}
    \begin{scope}
      \clip (\xmin,\ymin) rectangle (\xmax,\ymax);
      % --- regions -------------------------------------------------------
      % background: white (complex zeros, |x|<|y|)
      \fill[rW] (\xmin,\ymin) rectangle (\xmax,\ymax);
      % grey: complex zeros, |x|>|y|
      \fill[rG] (0.5,0.5) -- (10,10) -- (10,0.025)
        -- plot[domain=10:0.5,samples=200,smooth] (\x,{1/(4*\x)}) -- cycle;
      \fill[rG] (-0.5,-0.5) -- (-10,-10) -- (-10,-0.025)
        -- plot[domain=-10:-0.5,samples=200,smooth] (\x,{1/(4*\x)}) -- cycle;
      % red strip: |x+y|<1
      \fill[rR] (-10,11) -- (10,-9) -- (10,-11) -- (-10,9) -- cycle;
      % green: real zeros outside the strip, |x|<|y|
      \fill[rGr] (0.5,0.5) -- (-9,10) -- (0.025,10)
        -- plot[domain=10:0.5,samples=200,smooth] ({1/(4*\x)},\x) -- cycle;
      \fill[rGr] (-0.5,-0.5) -- (9,-10) -- (-0.025,-10)
        -- plot[domain=-10:-0.5,samples=200,smooth] ({1/(4*\x)},\x) -- cycle;
      % blue: real zeros outside the strip, |x|>|y|
      \fill[rB] (0.5,0.5)
        -- plot[domain=0.5:10,samples=200,smooth] (\x,{1/(4*\x)})
        -- (10,-9) -- cycle;
      \fill[rB] (-0.5,-0.5)
        -- plot[domain=-0.5:-10,samples=200,smooth] (\x,{1/(4*\x)})
        -- (-10,9) -- cycle;
      % --- boundaries ----------------------------------------------------
      \draw[bdry] (-10,11) -- (10,-9);                 % x+y=1
      \draw[bdry] (-10,9) -- (10,-11);                 % x+y=-1
      \draw[bdry] (0.5,0.5) -- (10,10);                % y=x, x>1/2
      \draw[bdry] (-0.5,-0.5) -- (-10,-10);            % y=x, x<-1/2
      \draw[bdry] plot[domain=0.025:10,samples=300,smooth] (\x,{1/(4*\x)});
      \draw[bdry] plot[domain=-10:-0.025,samples=300,smooth] (\x,{1/(4*\x)});
      % --- axes ------------------------------------------------------------
      % horizontal axis: remaining zero z=-t1/t2, inside D for |x|<1
      % vertical axis:   remaining zero z=-t2/t3, inside D for |y|>1
      \draw[halo] (\xmin,0) -- (\xmax,0);
      \draw[halo] (0,\ymin) -- (0,\ymax);
      % the dots stop a little away from +-1, leaving a visible gap
      \draw[axOut] (0,-0.9) -- (0,0.9);
      \draw[axIn]  (0,1) -- (0,\ymax);
      \draw[axIn]  (0,-1) -- (0,\ymin);
      \draw[axIn]  (-1,0) -- (1,0);
      \draw[axOut] (1.1,0) -- (\xmax,0);
      \draw[axOut] (-1.1,0) -- (\xmin,0);
    \end{scope}
    % --- ticks and labels ----------------------------------------------
    \foreach \t in {-2,2,4}
      \draw[line width=0.4pt] (\t,-0.1) -- (\t,0.1) node[below=3pt,fill=white,inner sep=1pt] {\small $\t$};
    \foreach \t in {-2,2}
      \draw[line width=0.4pt] (-0.1,\t) -- (0.1,\t) node[left=5pt,fill=white,inner sep=1pt] {\small $\t$};
    \node[below left=1pt,fill=white,inner sep=1pt] at (-0.05,-0.15) {\small $0$};
    \node[above left,fill=white,inner sep=1pt] at (\xmax,0.12) {$t_1/t_2$};
    \node[below right,fill=white,inner sep=1pt] at (0.08,\ymax) {$t_3/t_2$};
    \draw[line width=0.4pt] (\xmin,\ymin) rectangle (\xmax,\ymax);
    % --- legend ----------------------------------------------------------
    \begin{scope}[shift={(\xmin,\ymin-0.55)},
        every node/.style={font=\footnotesize,anchor=west,align=left,inner sep=1pt}]
      \def\sw#1#2{\draw[#1,line width=0.4pt] (#2) ++(0,-0.15) rectangle ++(0.3,0.3);}
      % left column
      \sw{rW}{0,0}      \node at (0.45,0)    {$2$ e.p., $|z|<1$, $z\notin\mathbb{R}$};
      \sw{rG}{0,-0.55}  \node at (0.45,-0.55) {$2$ e.p., $|z|>1$, $z\notin\mathbb{R}$};
      \sw{rB}{0,-1.1}   \node at (0.45,-1.1) {$2$ e.p., $|z|>1$, $z\in\mathbb{R}$};
      \draw[axOut] (0.03,-1.65) -- (0.3,-1.65);
                        \node at (0.45,-1.65) {$1$ e.p., $|z|>1$, $z\in\mathbb{R}$};
      % right column
      \sw{rGr}{4.3,0}   \node at (4.75,0)    {$2$ e.p., $|z|<1$, $z\in\mathbb{R}$};
      \sw{rR}{4.3,-0.825}\node at (4.75,-0.825) {$1$ e.p., $|z|>1$, $z\in\mathbb{R}$\\
            $+\,1$ e.p., $|z|<1$, $z\in\mathbb{R}$};
      \draw[axIn] (4.3,-1.65) -- (4.6,-1.65);
            \node at (4.75,-1.65) {$1$ e.p., $|z|<1$, $z\in\mathbb{R}$};
    \end{scope}
  \end{tikzpicture}
  \caption{Phase diagram describing the number of exceptional points at which $B=0$. In the legend, e.p.\ stands for exceptional point. On the horizontal axis ($t_3=0$), one exceptional point goes to infinity, leaving a single exceptional point at finite $z$; this limit is the standard SSH model. On the vertical axis ($t_1=0$), one exceptional point goes to $z=0$; the corresponding state is localized entirely on the first site, which is decoupled from the rest of the chain. The remaining exceptional point is that of a standard SSH model with hoppings $t_2$ and $t_3$.}
  \label{fig:phase diagram}
\end{figure}

Let us focus on the boundary given by case 1. We can compare our approach with using the Zak phase. The Zak phase in this model~\cite{Asboth} is given by
\begin{align*}
    \nu_Z&=\frac{1}{2\pi i}\int_{-\pi/a}^{\pi/a}dk \frac{d}{dk} \ln\Big( t_1+t_2e^{ika}+t_3 e^{2ika}\Big)\\
    &=\frac{1}{2\pi i}\oint_{S(1)} dz\,\partial_z \ln\Big( t_1+t_2z+t_3 z^{2}\Big).
\end{align*}
This integral is equal to the number of zeros of the $B$-site in $D$, which is $2\nu_B$. Therefore, for this choice of unit cell,
\[\nu_Z = 2\nu_B,\]
which is the correct number of adiabatically protected edge states. However, contrary to the Zak phase, our expression is independent of the choice of unit cell. Choosing the other unit cell ($B-A$ rather than $A-B$) is equivalent to $t_1\to t_2$, $t_2 \to t_1$, $t_3z^{\pm2} \to t_3z^{\mp 1}$ and $B\to A=B'$, $A\to B=A'$. The new value of $\nu_Z$ is
\[\nu_Z' = \frac{1}{2\pi i}\oint_{S(1)}dz\, \partial_z \ln (t_2 + t_1 z + t_3 z^{-1}),\]
which counts $-1 + (\# \text{ of zeros of }t_1 + t_2 z^{-1} + t_3 z^{-2} \text{ in }D)$, thus $\nu_Z' = -1 + 2\nu_A$. Noting that $\nu_A+\nu_B=1$ leads to \[\nu_Z'=1-2\nu_B\neq \nu_Z.\] However, the new value of $\nu_B$ is
\[\nu_B' = \frac{1}{4\pi i} \oint_{\partial D}dz\, \partial_z \ln(t_2 + t_1z^{-1}+t_3z),\]
which counts $\nu_B'=\frac{1}{2}(\# \text{ of zeros of }t_1+t_2z + t_3z^2 \text{ in }D )= \nu_B$, and is still (half) the correct number of edge states.

{\color{red}%
Finally, for boundary 1 this model can have two adiabatically protected zero-energy edge states at the same time (when $2\nu_B=2$). Both reside on the same sublattice, $A$. Since a chiral-symmetric perturbation anticommutes with $\Gamma$, it has no matrix elements between states of the same chirality, so these states cannot hybridize and move away from zero energy; this is the usual statement that the difference between the numbers of zero-energy edge states on the two sublattices is protected by chiral symmetry. In the present picture, both states are exceptional states of $h(z)$ with $|z|<1$, and they can only disappear from the semi-infinite chain by crossing the unit circle, which requires the bulk gap to close.
}%end red

\section{Conclusion}
\label{sec:conclusion}

{\color{red}%
We have provided a physically insightful derivation of the bulk-boundary correspondence in semi-infinite chiral chains, with arbitrary unit cells and hoppings of arbitrary range, which relies on complex analysis and linear algebra rather than on $K$-theory or index theorems. The important insights are that, when a single boundary condition suffices, protected edge states are analytically continued Bloch states with a sublattice zero, and that in chiral chains such states occur at the exceptional points of the analytically continued Bloch Hamiltonian. For longer-range hoppings, the exceptional states give a lower bound on the number of protected edge states. This latter insight also allows for a new interpretation of the  well-established winding number for chirally symmetric Hamiltonians.
}%end red

Note that, although one can use the bulk Bloch Hamiltonian to determine the number of edge states in a semi-infinite chain, it is perhaps an overstatement to describe this as a \emph{bulk-boundary correspondence}, in that the \emph{bulk} part of this relationship is not fully independent of the boundary. For instance, for the standard SSH model there are two possible unit cells ($AB$ or $BA$), and the bulk calculation -- the Zak phase -- will give different results for the two choices. The correct choice of unit cell is determined by the precise edge of the chain.

\begin{acknowledgments}
We thank Hichem Eleuch, Michael Hilke and Andrew Mckenna for useful conversations. This work was supported in part by the Natural Sciences and Engineering Research Council of Canada and by the Fonds de Recherche du Qu{\'e}bec -- Nature et Technologies  via the INTRIQ strategic cluster grant.
\end{acknowledgments}

\appendix

\section{Boundary conditions of semi-infinite chains}
\label{app:boundarySemiInfinite}

Much of what has been said previously relies on the assumption that the boundary condition $\langle p_b,\mu_b | \Psi \rangle = 0$ is sufficient for a semi-infinite system. Let us examine in more detail the possible boundary conditions of a semi-infinite chain in order to understand the edge states that can be described by our construction.

Let $\Hinf$ be the Hamiltonian of an infinite chain; suppose we are interested in the semi-infinite chain obtained by excising all sites to the left of a given site. It is useful to write $\Hinf$ as a $2\times2$ matrix of infinite-dimensional matrices:
% some tricks to get columns of equal width (\matrixcolwidth defined in preamble)
\begin{equation}
\renewcommand{\arraystretch}{1.3} 
\Hinf =
\settowidth{\matrixcolwidth}{$\Hsi$}%use widest matrix element
\begin{pmatrix}
    \begin{array}{w{c}
        { \matrixcolwidth} | w{c}{\matrixcolwidth}}
        V & W^\dagger\\
        \hline
        W & \Hsi 
    \end{array}
\end{pmatrix}.
\end{equation}
Here $V$ is the Hamiltonian of the excised semi-infinite chain, $\Hsi$ is the Hamiltonian of the system, and $W,W^\dagger$ represent the coupling between the two semi-infinite chains. The matrices $V$ and $\Hsi$ are of course Hermitian; the form of $W$ will be discussed presently.

We are looking for eigenstates of the infinite chain which give eigenstates of the system upon reduction. The Schrödinger equation for the infinite system can be written in block form:
\begin{equation}
    \Hinf\ket{\Psi} \equiv
    \Hinf \begin{pmatrix}
        \renewcommand{\arraystretch}{1.3}
        \begin{array}{@{}c@{}}
            \ket{\phi} \\
            \hline
            \ket{\psi}
        \end{array}
    \end{pmatrix}
    =
    \begin{pmatrix}
        \renewcommand{\arraystretch}{1.3}
        \begin{array}{@{}c@{}}
            V \ket{\phi} + W^\dagger \ket{\psi} \\
            \hline
            W \ket{\phi} + \Hsi\ket{\psi}
        \end{array}
    \end{pmatrix}
    =E
    \begin{pmatrix}
        \renewcommand{\arraystretch}{1.3}
        \begin{array}{@{}c@{}}
            \ket{\phi} \\
            \hline
            \ket{\psi}
        \end{array}
    \end{pmatrix},
\end{equation}
where $\ket{\Psi}$ is the wave function of the infinite chain and $\ket{\phi}$ and $\ket{\psi}$ are the parts corresponding to the excised semi-infinite chain and the system, respectively. From the second component of this equation, we see that $\ket{\psi}$ is a solution of the system's Schrödinger equation $\Hsi \ket{\psi} = E \ket{\psi}$ if and only if $W \ket{\phi} = 0$.

For a system with hopping range $R$ (defined in \autoref{sec:Bloch}), $W$ is of the form
\[
W =
\begin{pmatrix}
    % some tricks to get columns of equal width (\matrixcolwidth defined in preamble)
    \renewcommand{\arraystretch}{1.3} 
    \settowidth{\matrixcolwidth}{$W_R$}%use widest matrix element
    \begin{array}{w{c}{\matrixcolwidth} | w{c}{\matrixcolwidth}}
        0 & W_R \\
        \hline
        0 & 0
    \end{array}
\end{pmatrix}
\]
where $W_R$ is an $(R\times R)$-dimensional upper triangular matrix:
\begin{equation}
    W_R =\begin{pmatrix}
        w_{1,R} & w_{1,R-1} & ... & w_{1,2} & w_{1,1} \\
        0 & w_{2,R-1} & ... & w_{2,2} & w_{2, 1} \\
        \vdots & \ddots & \ddots & \vdots \\
        \vdots & & \ddots & w_{R-1,2} & w_{R-1, 1} \\   
        0 & \dots &  & 0 & w_{R, 1}    
    \end{pmatrix}.
\end{equation}
Labelling the matrix elements by their position relative to the top right corner is useful since then $w_{i,j}$ is the hopping linking the $i^\text{th}$ site of the semi-infinite system to the $j^\text{th}$ excised site. Of course, if not all hoppings up to range $R$ are present, some matrix elements within the triangle will be zero.

Under what circumstances is $W\ket{\phi}=0$? Letting $\ket{\phi_R}$ be the lower $R$ components of $\ket{\phi}$ (corresponding to the $R$ sites closest to the boundary), we see that $W\ket{\phi}=0$ if and only if
\[
W_R\ket{\phi_R}=0.
\]
This equation encapsulates the boundary conditions on the semi-infinite system. Thus, the number of boundary conditions is given by $N_\text{b} \equiv \text{rank}(W_R)$. For instance, if all diagonal elements are nonzero then $N_\text{b} = R$ and an eigenstate of the infinite chain truncates to an eigenstate of the semi-infinite chain if and only if $\ket{\phi_R}=0$.

For each row of zeros in $W_R$, there is one less boundary condition; therefore $N_\text{b} \leq  R$. Note that $N_\text{b}$ may depend not only on the length of the hoppings in a system, but also on the specific boundary under consideration.

We are now in a position to state under which circumstances our boundary condition $\langle p_b,\mu_b | \Psi \rangle =0$ is sufficient. This is true if only the rightmost column of $W_R$ is nonzero. The most natural way for this to occur is if $R=1$ (corresponding to a chain with only nearest-neighbour hoppings). But it is also possible for $R>1$ if the only hoppings that link sites of the system to excised sites land on the first excised site, $\ket{p_b,\mu_b}$.

Knowing that states of the analytically continued Bloch Hamiltonian which arise at exceptional points and have chiral symmetry are adiabatically protected, it is reasonable to assume that any topological edge state of a semi-infinite chain must arise as a linear combination of those exceptional states. Consider a chain with at least nearest-neighbour hoppings and chiral symmetry. Its chiral symmetry operator is necessarily 
\begin{equation}
    \Gamma = \begin{pmatrix}
        \ddots \\
        & 1 \\
        & & -1 \\
        & & & 1\\
        & & & & -1 \\
        & & & & & \ddots
    \end{pmatrix}.
\end{equation}
Edge states protected by chiral symmetry must therefore have the form
\begin{equation}
    |\psi \rangle = 
    \begin{pmatrix}
        \vdots\\
        \psi_{2i}\\
        0\\
        \psi_{2i+2} \\
        0\\ 
        \vdots
    \end{pmatrix} \quad\text{ or }\quad
    \begin{pmatrix}
        \vdots\\
        0\\
        \psi_{2i+1}\\
        0 \\
        \psi_{2i+3}\\ 
        \vdots
    \end{pmatrix}.
\end{equation}
One of the states above always satisfies at least half of the $N_\text{b}$ conditions; therefore, in reality there are at most $\lfloor R/2 \rfloor$ conditions left to satisfy for the exceptional states. {\color{red}%
Let us define $N_\text{b}'$ as the number of boundary conditions left to satisfy in the subspace of exceptional states. If a system has $K$ exceptional states, then in the space of exceptional states, the set of boundary conditions leads to $N_\text{b}'$ linear equations with $K$ variables, which
(assuming the equations are linearly independent, which is generically the case)
have $K-N_\text{b}'$ independent solutions. If a system has $K\leq N_\text{b}'$ exceptional states, it cannot have any adiabatically protected edge states.
}%end red

{\color{red}%
\section{Example: four-band model}
\label{app:4band}

As an example of a multiband model with several boundary conditions, we consider the chain shown in \autoref{fig:4band}. Its unit cell contains four sites, which we label $A_1$, $A_2$, $B_1$, $B_2$ for reasons that will become clear shortly, and the (real) hopping parameters $t_1,\dots,t_4$ are indicated in the figure. As shown in the bottom part of the figure, the chain is equivalent to a two-leg ladder. The model is simple enough to be solved analytically, but, unlike the extended SSH model of \autoref{sec:extendedSSH}, it has a genuinely matrix-valued $Q(z)$ and, for some choices of boundary (that is, of the site at which the semi-infinite chain begins), more than one boundary condition to be satisfied by the exceptional states.

\begin{figure}
  \centering
  \resizebox{\linewidth}{!}{%
  \begin{tikzpicture}[
      site/.style={circle,fill=black,inner sep=1.4pt},
      lab/.style={font=\footnotesize},
      tone/.style={line width=1.3pt},
      ttwo/.style={line width=0.5pt},
      tthree/.style={line width=1.3pt,dash pattern=on 3pt off 2pt},
      tfour/.style={line width=0.5pt,dash pattern=on 3pt off 2pt}]
    \def\dx{0.7}
    % ---------- 1D chain, hoppings t1, t2 ----------
    \begin{scope}[yshift=0cm]
      \foreach \k in {0,...,11}{
        \pgfmathtruncatemacro{\m}{mod(\k,4)}
        \node[site] (a\k) at (\k*\dx,0) {};
        \ifcase\m
          \node[lab,below=1pt] at (a\k) {$A_1$};\or
          \node[lab,above=1pt] at (a\k) {$A_2$};\or
          \node[lab,below=1pt] at (a\k) {$B_1$};\or
          \node[lab,above=1pt] at (a\k) {$B_2$};\fi}
      % A1-B1 and A2-B2 (t1, within cell); B1-A1' and B2-A2' (t2, between cells)
      \foreach \k in {0,4,8}{
        \pgfmathtruncatemacro{\kb}{\k+2}
        \draw[tone] (a\k) to[out=80,in=100,looseness=1.3] (a\kb);}
      \foreach \k in {1,5,9}{
        \pgfmathtruncatemacro{\kb}{\k+2}
        \draw[tone] (a\k) to[out=-80,in=-100,looseness=1.3] (a\kb);}
      \foreach \k in {2,6}{
        \pgfmathtruncatemacro{\kb}{\k+2}
        \draw[ttwo] (a\k) to[out=80,in=100,looseness=1.3] (a\kb);}
      \foreach \k in {3,7}{
        \pgfmathtruncatemacro{\kb}{\k+2}
        \draw[ttwo] (a\k) to[out=-80,in=-100,looseness=1.3] (a\kb);}
    \end{scope}
    \node at (5.5*\dx,-1.1) {$+$};
    % ---------- 1D chain, hoppings t3, t4 ----------
    \begin{scope}[yshift=-2.2cm]
      \foreach \k in {0,...,11}{
        \pgfmathtruncatemacro{\m}{mod(\k,4)}
        \node[site] (b\k) at (\k*\dx,0) {};
        \ifcase\m
          \node[lab,below=1pt] at (b\k) {$A_1$};\or
          \node[lab,above=1pt] at (b\k) {$A_2$};\or
          \node[lab,anchor=north east,inner sep=1pt] at (b\k) {$B_1$};\or
          \node[lab,anchor=south west,inner sep=1pt] at (b\k) {$B_2$};\fi}
      % A2-B1 (t3, within cell) and B2-A1' (t4, between cells): neighbours
      \foreach \k in {1,5,9}{
        \pgfmathtruncatemacro{\kb}{\k+1}
        \draw[tthree] (b\k) -- (b\kb);}
      \foreach \k in {3,7}{
        \pgfmathtruncatemacro{\kb}{\k+1}
        \draw[tfour] (b\k) -- (b\kb);}
      % A1-B2 (t3, within cell) and B1-A2' (t4, between cells): range 3
      \foreach \k in {0,4,8}{
        \pgfmathtruncatemacro{\kb}{\k+3}
        \draw[tthree] (b\k) to[out=75,in=105,looseness=1.25] (b\kb);}
      \foreach \k in {2,6}{
        \pgfmathtruncatemacro{\kb}{\k+3}
        \draw[tfour] (b\k) to[out=-75,in=-105,looseness=1.25] (b\kb);}
    \end{scope}
    \node at (5.5*\dx,-3.35) {$=$};
    % ---------- ladder ----------
    \begin{scope}[xshift=.35cm,yshift=-4.1cm]
      \def\lx{1.4}
      \foreach \c in {0,...,5}{
        \pgfmathtruncatemacro{\parity}{mod(\c,2)}
        \node[site] (t\c) at (\c*\lx,0) {};
        \node[site] (u\c) at (\c*\lx,-1.1) {};
        \ifnum\parity=0
          \node[lab,above=1pt] at (t\c) {$A_1$};
          \node[lab,below=1pt] at (u\c) {$A_2$};
        \else
          \node[lab,above=1pt] at (t\c) {$B_1$};
          \node[lab,below=1pt] at (u\c) {$B_2$};
        \fi}
      \foreach \c in {0,2,4}{
        \pgfmathtruncatemacro{\d}{\c+1}
        \draw[tone]   (t\c) -- (t\d);  \draw[tone]   (u\c) -- (u\d);
        \draw[tthree] (t\c) -- (u\d);  \draw[tthree] (u\c) -- (t\d);}
      \foreach \c in {1,3}{
        \pgfmathtruncatemacro{\d}{\c+1}
        \draw[ttwo]  (t\c) -- (t\d);  \draw[ttwo]  (u\c) -- (u\d);
        \draw[tfour] (t\c) -- (u\d);  \draw[tfour] (u\c) -- (t\d);}
      % unit cell
      \draw[gray,rounded corners=3pt,line width=0.4pt]
        (2*\lx-0.35,0.45) rectangle (3*\lx+0.35,-1.55);
    \end{scope}
    % ---------- legend ----------
    \begin{scope}[xshift=8.4cm,yshift=-0.6cm]
      \draw[tone]   (0,0)    -- (0.6,0);    \node[right] at (0.6,0)    {$t_1$};
      \draw[ttwo]   (0,-0.5) -- (0.6,-0.5); \node[right] at (0.6,-0.5) {$t_2$};
      \draw[tthree] (0,-1.0) -- (0.6,-1.0); \node[right] at (0.6,-1.0) {$t_3$};
      \draw[tfour]  (0,-1.5) -- (0.6,-1.5); \node[right] at (0.6,-1.5) {$t_4$};
    \end{scope}
  \end{tikzpicture}}
  \caption{Four-band model. The chain (top two lines, showing the hoppings $t_1,t_2$ and $t_3,t_4$ separately) is equivalent to the two-leg ladder shown in the bottom line. Thick lines denote hoppings within a unit cell ($t_1$, $t_3$) and thin lines hoppings between neighbouring unit cells ($t_2$, $t_4$). The box indicates a unit cell.}
  \label{fig:4band}
\end{figure}

In the basis $(A_1,A_2,B_1,B_2)$, the analytically continued Bloch Hamiltonian is
\begin{align}\label{eq:4band h}
    h(z) &=
    \begin{pmatrix}
    % some tricks to get columns of equal width (\matrixcolwidth defined in preamble)
        \renewcommand{\arraystretch}{1.3}
        \settowidth{\matrixcolwidth}{$Q(1/z^*)^\dagger$}%use widest matrix element
        \begin{array}{w{c}{\matrixcolwidth} | w{c}{\matrixcolwidth}}
            0 & Q(z)\\
            \hline
            Q(1/z^*)^\dagger & 0
        \end{array}
    \end{pmatrix},
    \\
    Q(z) &=
    \begin{pmatrix}
        t_1 + t_2z^{-1} & t_3 + t_4 z^{-1}\\
        t_3 + t_4 z^{-1} & t_1 + t_2z^{-1}
    \end{pmatrix}\nonumber\\
    &= (t_1+t_2z^{-1})\mathds{1}+(t_3+t_4z^{-1})\sigma_x.\label{eq:Qform}
\end{align}
It has chiral symmetry, $\{\Gamma,h(z)\}=0$, and a reflection symmetry, $[R,h(z)]=0$, where
\begin{equation}\label{eq:4band symmetries}
    \Gamma =
    \begin{pmatrix}
        \renewcommand{\arraystretch}{1.3}
    % some tricks to get columns of equal width (\matrixcolwidth defined in preamble)
        \settowidth{\matrixcolwidth}{$-\mathds{1}$}%use widest matrix element
        \begin{array}{w{c}{\matrixcolwidth} | w{c}{\matrixcolwidth}}
            \mathds{1} & 0\\
            \hline
            0 & -\mathds{1}
        \end{array}
    \end{pmatrix},
    \qquad
    R =
     \begin{pmatrix}
        \renewcommand{\arraystretch}{1.3}
        \settowidth{\matrixcolwidth}{$\sigma_x$}%use widest matrix element
        \begin{array}{w{c}{\matrixcolwidth} | w{c}{\matrixcolwidth}}
            \sigma_x & 0\\
            \hline
            0 & \sigma_x
        \end{array}
    \end{pmatrix}.
\end{equation}
Thus $h(z)$ is already in the form \eqref{eq:chiral hamiltonian} of \autoref{sec:chiralSymmetry}, with sublattice $A=\{A_1,A_2\}$ and sublattice $B=\{B_1,B_2\}$. The reflection $R$ exchanges $A_1\leftrightarrow A_2$ and $B_1\leftrightarrow B_2$, that is, the two legs of the ladder are interchanged. Given the form of
$Q(z)$ (see \eqref{eq:Qform}), the condition $[R,h(z)]=0$ is equivalent to $[\sigma_x,Q(z)]=0$.

Since $[R,\Gamma]=0$, the two symmetries can be diagonalized simultaneously. In the sectors $R=\pm1$, spanned by $(A_1\pm A_2)/\sqrt{2}$ and $(B_1\pm B_2)/\sqrt{2}$, the Bloch Hamiltonian reduces to
\begin{equation}\label{eq:4band hpm}
    h_{\pm}(z) =
    \begin{pmatrix}
         0 & f_\pm(z)\\
         f_\pm(1/z) & 0
    \end{pmatrix},
\end{equation}
where
\begin{equation}
    f_\pm(z) = t_1 \pm t_3 + (t_2 \pm t_4)z^{-1},
\end{equation}each of which is the Bloch Hamiltonian of a standard SSH chain with intracell hopping $t_1\pm t_3$ and intercell hopping $t_2\pm t_4$. In particular, $\det Q(z)=f_+(z)f_-(z)$ and $\det Q(1/z^*)^\dagger=f_+(1/z) f_-(1/z)$. The exceptional points with $A_1=A_2=0$ are therefore the zeros of $\det Q(z)$, at $z=-(t_2\pm t_4)/(t_1\pm t_3)$, and those with $B_1=B_2=0$ are the zeros of $\det Q(1/z^*)^\dagger$, at $z=-(t_1\pm t_3)/(t_2\pm t_4)$. The determinant of $Q(z)$ has a pole of order two at the origin.

The eigenstates of $h(z)$ are
\begin{equation}\label{eq:4band states}
    \psi_{s_1,s_2}(z) =
    \begin{pmatrix}
        \sqrt{f_{s_1}(z)}\\
        s_1\sqrt{f_{s_1}(z)}\\
        s_2\sqrt{f_{s_1}(1/z)}\\
        s_1s_2\sqrt{f_{s_1}(1/z)}
    \end{pmatrix},
    \qquad s_1,s_2=\pm1,
\end{equation}
with energy $E=s_2\sqrt{f_{s_1}(z)f_{s_1}(1/z)}$, where $s_1$ is the reflection eigenvalue and $s_2$ the sign of the energy within the corresponding sector. It follows that
\begin{equation}\label{eq:4band UA}
    U_A(z) =
    \begin{pmatrix}
        \sqrt{f_+(z)} & \sqrt{f_-(z)}\\
        \sqrt{f_+(z)} & -\sqrt{f_-(z)}
    \end{pmatrix},
\end{equation}
and $U_B(z)=U_A(1/z)$. Thus $\det U_A(z)=-2\sqrt{\det Q(z)}$, which vanishes as a square root at the exceptional points with $A_1=A_2=0$ and has a simple pole at the origin. Going around an exceptional point with $A_1=A_2=0$, the column of $U_A(z)$ belonging to the corresponding sector changes sign, and with it $\det U_A(z)$; around an exceptional point with $B_1=B_2=0$, $U_A(z)$ is unchanged. In either case $\partial_z\ln\det U_A(z)$ is single-valued, and
\begin{align}\label{eq:4band nuA}
    2\nu_A &= \frac{1}{\pi i}\oint_{\partial D} dz\, \partial_z \ln \det U_A(z)\nonumber\\
    &= \frac{1}{2\pi i}\oint_{\partial D} dz\, \partial_z \ln \det Q(z)
\end{align}
is the number of exceptional states with $A_1=A_2=0$ in $D$, that is, the number of sectors with $|t_2\pm t_4|<|t_1\pm t_3|$. Similarly,
\begin{align}\label{eq:4band nuB}
    2\nu_B &= \frac{1}{\pi i}\oint_{\partial D} dz\, \partial_z \ln \det U_B(z)\nonumber\\
    &= \frac{1}{2\pi i}\oint_{\partial D} dz\, \partial_z \ln \det Q(1/z^*)^\dagger
\end{align}
is the number of exceptional states with $B_1=B_2=0$ in $D$, that is, the number of sectors with $|t_1\pm t_3|<|t_2\pm t_4|$. Generically, $2\nu_A+2\nu_B=2$.

We now consider a semi-infinite chain with the sites ordered as in the top part of \autoref{fig:4band}, $\dots,A_1,A_2,B_1,B_2,A_1,\dots$. Of the four possible boundaries, those in which the chain begins with $B_1$ or $B_2$ are obtained from those in which it begins with $A_1$ or $A_2$ by shifting the unit cell by two sites, which exchanges the roles of the two sublattices and of the hoppings ($t_1\leftrightarrow t_2$, $t_3\leftrightarrow t_4$). It is therefore sufficient to consider the following two boundaries:
\begin{enumerate}
    \item The boundary is between $B_2$ and $A_1$ (the chain begins with a complete unit cell).
    \item The boundary is between $A_1$ and $A_2$ (the chain begins with $A_2$).
\end{enumerate}
We denote by $A_{1b}$, $A_{2b}$, $B_{1b}$ and $B_{2b}$ the amplitudes on the last excised site of each type. 

For boundary 1, only $B_{1b}$ and $B_{2b}$ are coupled to the system, and the boundary conditions are $t_2B_{1b}+t_4B_{2b}=0$ and $t_4B_{1b}+t_2B_{2b}=0$, that is (for $t_2\neq\pm t_4$), $B_{1b}=B_{2b}=0$, so that $N_\text{b}=2$. Exceptional states with $B_1=B_2=0$ satisfy both conditions automatically ($N_\text{b}'=0$), so that each of them is an edge state, while exceptional states with $A_1=A_2=0$ would have to satisfy both ($N_\text{b}'=2$), which is impossible since $2\nu_A\leq2$. This boundary preserves the reflection symmetry, so that each sector is a semi-infinite SSH chain with the usual termination, and the result, $2\nu_B$ edge states on sublattice $A$, is simply the number of sectors in the topological phase.

For boundary 2, the excised site $A_1$ of the last unit cell is coupled to $B_1$ and $B_2$ of the first unit cell of the system, while $B_{1b}$ and $B_{2b}$ are coupled to $A_2$. The boundary conditions are $A_{1b}=0$ and $t_4B_{1b}+t_2B_{2b}=0$, so that again $N_\text{b}=2$. Exceptional states with $B_1=B_2=0$ must satisfy the first condition and those with $A_1=A_2=0$ the second, so that $N_\text{b}'=1$ in both cases; from \eqref{eq:edge-state-count}, there are at least $2\nu_B-1$ edge states on sublattice $A$ and at least $2\nu_A-1$ on sublattice $B$. The resulting numbers of edge states are given in \autoref{tab:4band}.

In each case, removing the single site $A_1$ (going from boundary 1 to boundary 2) changes the difference between the numbers of edge states on sublattices $A$ and $B$ by $-1$, as expected from chiral symmetry, since it reduces the number of $A$ sites by one.

We have checked these results by exact diagonalization of finite chains of $L=60$ unit cells with both boundaries on the left and all four possible boundaries on the right, for several sets of hopping parameters in each of the three phases of \autoref{tab:4band}. In all cases, the number of zero modes localized at the left end of the chain agrees with \autoref{tab:4band}. In a finite chain, an edge state at the left end can hybridize with one at the right, and the pair is split by an energy that decays as $|z_\text{e.p.}|^L$, where $z_\text{e.p.}$ is the corresponding exceptional point. We have verified this exponential decay, including its rate, which confirms that the near-zero modes are edge states of the semi-infinite chain, split only by the finite length.

\begin{table}
  \caption{Number of zero-energy edge states of the four-band model for the two boundaries considered in the text, with the sublattice on which they reside. Here $2\nu_B$ ($2\nu_A$) is the number of sectors in which $|t_1\pm t_3|<|t_2\pm t_4|$ ($|t_1\pm t_3|>|t_2\pm t_4|$).}
  \label{tab:4band}
  \begin{ruledtabular}
  \begin{tabular}{cccc}
    $2\nu_A$ & $2\nu_B$ & Boundary 1 & Boundary 2\\
    \hline
    0 & 2 & 2 ($A$) & 1 ($A$)\\
    1 & 1 & 1 ($A$) & 0\\
    2 & 0 & 0 & 1 ($B$)\\
  \end{tabular}
  \end{ruledtabular}
\end{table}

}%end red

\bibliography{ref}

\end{document}